\documentclass[a4paper,fleqn]{cas-dc}
\usepackage[numbers,sort&compress]{natbib}
\usepackage{subcaption}
\usepackage{threeparttable}
\usepackage{latexsym} 
\usepackage{pdflscape} 
\usepackage{afterpage} 

\usepackage[T1]{fontenc} 
\usepackage[utf8]{inputenc}
\usepackage[english]{babel}

\usepackage[most]{tcolorbox}
\tcbuselibrary{theorems}
\usepackage{cleveref}
\newtcbtheorem[no counter]{Results}{Summary for}{
  enhanced,
  unbreakable,
  sharp corners,
  attach boxed title to top left={
    yshifttext=0mm
  },
  colback=white,
  colframe=black,
  coltitle=black,
  fonttitle=\bfseries,
  boxed title style={
    sharp corners,
    size=small,
    colback=white,
    colframe=black,
  },
  separator sign none
}{summary}

\graphicspath{{figs/}}

\begin{document}
\let\WriteBookmarks\relax
\def\floatpagepagefraction{1}
\def\textpagefraction{.001}

\shorttitle{Are Game Engines Software Frameworks? A Three-perspective Study}
\shortauthors{Politowski et~al.}

\title[mode=title]{Are Game Engines Software Frameworks? A Three-perspective Study}
\tnotemark[1]

\tnotetext[1]{Dataset: \url{https://doi.org/10.5281/zenodo.3606899}.}


\author[1]{Cristiano Politowski}[orcid=0000-0002-0206-1056] \cormark[1]
\ead{c_polito@encs.concordia.ca}

\author[2]{Fabio Petrillo}[orcid=0000-0002-8355-1494]
\ead{fabio@petrillo.com}

\author[3]{João Eduardo Montandon}[orcid=0000-0002-3371-7353]
\ead{joao.montandon@dcc.ufmg.br}

\author[3]{Marco Tulio Valente}[orcid=0000-0002-8180-7548]
\ead{mtov@dcc.ufmg.br}

\author[1]{Yann-Ga\"el Gu\'{e}h\'{e}neuc}[orcid=0000-0002-4361-2563]
\ead{yann-gael.gueheneuc@concordia.ca}

\address[1]{Concordia University, Montreal, Quebec, Canada}
\address[2]{Université du Québec à Chicoutimi, Chicoutimi, Quebec, Canada}
\address[3]{Universidade Federal de Minas Gerais, Belo Horizonte, Brazil}

\cortext[cor1]{Corresponding author}

\newcommand{\projectpage}{https://doi.org/10.5281/zenodo.3606899}

\newtcolorbox{summarybox}{
    arc=0mm,
    colback=white,
    boxrule=0.1mm}

\newcommand{\todo}[1]{\textcolor{blue}{TODO: #1} }
\newcommand{\Fabio}[1]{\textcolor{red}{$>>>$ Fabio: #1 $<<<$}}
\newcommand{\YANN}[1]{\textcolor{red}{$>>>$ Yann: #1 $<<<$}}
\newcommand{\CRIS}[1]{\textcolor{red}{$>>>$ Cris: #1 $<<<$}}
\newcommand{\Joao}[1]{\textcolor{red}{$>>>$ Joao: #1 $<<<$}}



\begin{abstract}
Game engines help developers create video games and avoid duplication of code and effort, like frameworks for traditional software systems. In this paper, we explore open-source game engines along three perspectives: literature, code, and human. First, we explore and summarise the academic literature on game engines. Second, we compare the characteristics of the 282 most popular engines and the 282 most popular frameworks in GitHub. Finally, we survey 124 engine developers about their experience with the development of their engines. We report that: (1) Game engines are not well-studied in software-engineering research with few studies having engines as object of research. (2) Open-source game engines are slightly larger in terms of size and complexity and less popular and engaging than traditional frameworks. Their programming languages differ greatly from frameworks. Engine projects have shorter histories with less releases. (3) Developers perceive game engines as different from traditional frameworks. Generally, they build game engines to (a) better control the environment and source code, (b) learn about game engines, and (c) develop specific games. 
We conclude that open-source game engines have differences compared to traditional open-source frameworks although this differences do not demand special treatments. 
\end{abstract}

\maketitle


\section{Introduction}

\begin{quote}
``\textit{It's hard enough to make a game (...). It's really hard to make a game where you have to fight your own tool set all the time.}'' \newline--- \citet{SchreierKotaku2019} quoting a game developer on the difficulties faced using their game engine.
\end{quote}

For decades, video games have been a joyful hobby for many people around the world \cite{esa18}, making the game industry multi-billionaire, surpassing the movie and music industries combined \cite{Newzoo19}. However, realistic graphics and smooth gameplays hide constant and non-ending problems with game development, mostly related to poor software-development practices and inadequate management \cite{Petrillo2009}. Problems like these result in a scenario where 80\% of the top 50 games on Steam\footnote{Steam is a video game digital distribution service platform available at \url{https://store.steampowered.com/}.} need critical updates \cite{Lin2017:updates}, also leaving a trail of burnout developers after long periods of ``crunchs''\footnote{In game development, ``crunch time'' is the period during which developers work extra hours to deliver their game in time.} \cite{Edholm2017}.

During game development, developers use specialized software infrastructures to develop their games; chief among which are \textit{game engines}. Game engines encompass a myriad of resources and tools \cite{Gregory2014, Thorn2011, Hughes2010, Sherrod2007}. They can be built from scratch during game development, reused from previous games, extended from open-source ones, or bought off the shelves. They are essential to game development but misunderstood and misrepresented by the media \cite{SchreierKotaku:engine} and developers due to lacks of clear definitions, architectural references \cite{Messaoudi2016}, and academic studies. They are also the source of problems, especially between design and technical teams \cite{Schreier2017, Kushner2003}.


To address these problems, some researchers suggest the use of software-engineering techniques \cite{Petrillo2009, Petrillo2010, Kasurinen2017} while others consider game development as a special kind of software and propose new engineering practices or extensions to classical ones \cite{Hyrynsalmi2018, Murphy-Hill2014, Lewis2011, Kanode2009, Callele2015, Ramadan2014, Mozgovoy2018}. However, they did not study a large number of game engines, either proprietary, because only 13\% of all the games on Steam describe their engines \cite{Toftedahl2019}, or open source. They also did not survey game engine developers.

Therefore, we set to comparing open-source video-game engines with traditional open-source software frameworks can help researchers and developers to understand them better. With this article, we want to answer whether game engines share similar characteristics with software frameworks. By comparing the tools (engines and frameworks) rather than their instances (video games, traditional software systems), we provide a distinct view on game development: rather than studying how developers use games engines, we focus on how the foundations of their games are built.

We study open-source game engines from three perspectives: \emph{literature}, \emph{code}, and \emph{human} to provide an global view on the states of the art and practice on game engines. We explore academic and gray literature on game engines; compare the characteristics of the 282 most popular engines and the 282 most popular frameworks in GitHub; and, survey 124 engine developers about their experience with the development of their engines. 
Thus, we provide four contributions: (1) a corpus of open-source engines for this and future research work; (2) an analysis and discussion of the characteristics of engines; (3) a comparison of these characteristics with those of traditional frameworks; and, (4) a survey of engine developers about their experience with engine development.


We show that, different from what researchers and engine developers think, there are qualitative but no quantitative differences between engines and frameworks. 
Game engines are slightly larger in terms of size and complexity and less popular and engaging than traditional frameworks. The programming languages of game engines differ greatly from that of traditional frameworks. game engine projects have shorter histories with less releases. Developers perceive game engines as different from traditional frameworks and claim that engines need special treatments. Developers build game engines to better control the environment and source code, to learn about game engines, and to develop specific games. 
We conclude that open-source game engines have differences compared to traditional open-source frameworks although this differences, related to community, do not demand special treatments.

The paper is structured as follows: 
Section~\ref{sec:rqs} lists the research questions for all the three perspectives: literature, code, human. 
Section~\ref{sec:rq1-literature} shows the results of the literature perspective.
Section~\ref{sec:study-design} described the study design for the code perspective: metrics, data collection, and analysis.
Section~\ref{sec:disc} discusses our results and threats to their validity. 
Section~\ref{sec:conclusion} concludes.

In addition, the detailed results are in the Appendix Sections.
Appendix~\ref{sec:rq2-static} shows results related to static analysis of the projects.
Appendix~\ref{sec:rq3-historic} shows the historic analysis of the projects.
Appendix~\ref{sec:rq4-community} shows the community analysis of the projects.
Appendix~\ref{sec:rq5-human} describes our survey of engine developers for the human perspective.



\newcommand{\rqA}{RQ1: Literature Perspective}
\newcommand{\rqAa}{%
    RQ1.1: What is the definition for \textit{software framework}?
}
\newcommand{\rqAb}{%
    RQ1.2: What is the definition for \textit{game engine}?
}
\newcommand{\rqAc}{%
    RQ1.3: What are the \textit{works} related to game engines?
}


\newcommand{\rqBCD}{RQ2: Code Perspective}

\newcommand{\rqB}{RQ2.1: Static Characteristics}
\newcommand{\rqBa}{%
    RQ2.1.1: What is the popularity of the \emph{languages} in the projects?
}
\newcommand{\rqBb}{%
    RQ2.1.2: What is the popularity of the \emph{licenses} in the projects?
}
\newcommand{\rqBc}{%
    RQ2.1.3: What are the \emph{project sizes} of engines and frameworks?
}
\newcommand{\rqBd}{%
    RQ2.1.4: What are the \emph{function sizes} of engines and frameworks?
}
\newcommand{\rqBe}{%
    RQ2.1.5: What are the \emph{function complexities} of engines and frameworks?
}
\newcommand{\mBa}{%
    \texttt{main\_language}
}
\newcommand{\mBb}{%
    \texttt{license}
}
\newcommand{\mBcA}{%
    \texttt{main\_language\_size}
}
\newcommand{\mBcB}{%
    \texttt{total\_size}
}
\newcommand{\mBcC}{%
    \texttt{n\_files}
}
\newcommand{\mBdA}{%
    \texttt{n\_funcs}
}
\newcommand{\mBdB}{%
    \texttt{nloc\_mean}
}
\newcommand{\mBdC}{%
    \texttt{func\_per\_file\_mean}
}
\newcommand{\mBe}{%
    \texttt{cc\_mean}
}


\newcommand{\rqC}{RQ2.2: Historical Characteristics}

\newcommand{\rqCa}{%
    RQ2.2.1: How many versions were \emph{released} for each project?
}
\newcommand{\rqCb}{%
    RQ2.2.2: What is the \emph{lifetime} of the projects?
}
\newcommand{\rqCc}{%
    RQ2.2.3: How frequently do projects receive new \emph{contributions}?
}
\newcommand{\rqCd}{%
    RQ2.2.4: Are commits made on game engines more \emph{effort-prone}?
}
\newcommand{\mCa}{%
    \texttt{tags\_releases\_count}
}
\newcommand{\mCb}{%
    \texttt{lifespan}
}
\newcommand{\mCcA}{%
    \texttt{commits\_count}
}
\newcommand{\mCcB}{%
    \texttt{commits\_per\_time}
}
\newcommand{\mCdA}{%
    \texttt{lines\_added}
}
\newcommand{\mCdB}{%
    \texttt{lines\_removed}
}
\newcommand{\mCdC}{%
    \texttt{code\_churn}
}


\newcommand{\rqD}{RQ2.3: Community Characteristics}

\newcommand{\rqDa}{%
    RQ2.3.1: How many developers \emph{contribute} in the project?
}
\newcommand{\rqDb}{%
    RQ2.3.2: How \emph{popular} are the projects considering their main languages?
}
\newcommand{\rqDc}{%
    RQ2.3.3: How many \emph{issues} are reported in each project?
}
\newcommand{\mDa}{%
    \texttt{truck\_factor}
}
\newcommand{\mDbA}{%
    \texttt{stargazers\_count}
}
\newcommand{\mDbB}{%
    \texttt{contributors\_count}
}
\newcommand{\mDcA}{%
    \texttt{issues\_count}
}
\newcommand{\mDcB}{%
    \texttt{closed\_issues\_count}
}
\newcommand{\mDcC}{%
    \texttt{closed\_issues\_rate}
}


\newcommand{\rqE}{RQ3: Human Perspective}

\newcommand{\rqEa}{RQ3.1: What are the reasons developers create open-source game engines?}
\newcommand{\rqEb}{RQ3.2: Do game engine developers also have expertise with traditional software?}
\newcommand{\rqEc}{RQ3.3: For game engine developers, is it similar to developing a traditional framework?}

\newcommand{\mEa}{%
    Why did you create or collaborated with a video-game engine project?
}
\newcommand{\mEb}{%
    Have you ever written code for a software unrelated to games, like a Web, phone, or desktop app?
}
\newcommand{\mEc}{%
    How similar do you think writing a video-game engine is compared to writing a framework for traditional apps? (Like Django, Rails, or Vue)
}


\section{Research Questions}
\label{sec:rqs}

This Section shows the list of research questions and the metrics used to answer them. An overview of the Perspectives, RQs, and Metrics, is in the Appendix \autoref{fig:perspectives}. More details about the metrics are into the \autoref{tab:metrics}.

\subsection{\rqA}

Although software frameworks are part of the toolset of most developers nowadays, its concept is often misunderstood, specially with libraries\footnote{We encountered many problems while gathering the dataset as developers labeled their projects incorrectly; for example, labelling a library a ``framework'' or a game an ``engine''.}. To better understand the differences between frameworks and game engines we explore the \textit{literature} perspective using Scopus\footnote{\url{https://www.scopus.com}}, for academic books and the \textit{search engines} on internet, for articles, technical blogs, and discussion forums.
In the Section~\ref{sec:rq1-literature} we aim to answer the following research questions.

\begin{itemize}
    \item \rqAa 
    \item \rqAb 
    \item \rqAc 
\end{itemize}

\subsection{\rqBCD}

With respect to the design and implementation of game engines and traditional frameworks, we study their static, historical, and community characteristics.

\subsubsection*{\rqB}

To understand the differences of the frameworks and game engines projects from a code perspective, we investigate the static attributes of the projects, like their size, complexity of the functions, programming languages and licenses used. 
In the Section~\ref{sec-res-summary-rqb} (and Appendix~\ref{sec:rq2-static}) we aim to answer the following research questions.

\begin{itemize}
    \item \rqBa 
    \begin{itemize}
        \item Metrics: \mBa
    \end{itemize}
    \item \rqBb
    \begin{itemize}
        \item Metrics: \mBb
    \end{itemize}
    \item \rqBc
    \begin{itemize}
        \item Metrics: \mBcA, \mBcB, \mBcC
    \end{itemize}
    \item \rqBd
    \begin{itemize}
        \item Metrics: \mBdA, \mBdB, \mBdC
    \end{itemize}
    \item \rqBe
    \begin{itemize}
        \item Metrics: \mBe
    \end{itemize}
\end{itemize}

\subsubsection*{\rqC}

To explore the historical characteristics of the projects, we compare the life-cycles of game engines and traditional frameworks. We analyze the tags released (versions), projects' lifespan and commits.
In the Section~\ref{sec-res-summary-rqc} (and Appendix~\ref{sec:rq3-historic}) we aim to answer the following research questions.

\begin{itemize}
    \item \rqCa
    \begin{itemize}
        \item Metrics: \mCa
    \end{itemize}
    \item \rqCb
    \begin{itemize}
        \item Metrics: \mCb
    \end{itemize}
    \item \rqCc
    \begin{itemize}
        \item Metrics: \mCcA, \mCcB
    \end{itemize}
    \item \rqCd
    \begin{itemize}
        \item Metrics: \mCdA, \mCdB, \mCdC
    \end{itemize}
\end{itemize}

\subsubsection*{\rqD}

To investigate the interactions of the OSS community on the projects, we analyze the popularity of the projects, the number of issues reported in these projects, and the truck-factor measure~\cite{Avelino2016}.
In the Section~\ref{sec-res-summary-rqd} (and Appendix~\ref{sec:rq4-community}) we aim to answer the following research questions.

\begin{itemize}
    \item \rqDa
    \begin{itemize}
        \item Metrics: \mDa
    \end{itemize}
    \item \rqDb
    \begin{itemize}
        \item Metrics: \mDbA, \mDbB
    \end{itemize}
    \item \rqDc
    \begin{itemize}
        \item Metrics: \mDcA, \mDcB, \\\mDcC
    \end{itemize}
\end{itemize}

\subsection{\rqE}

The human perspective pertains to the developers' perception of game engines and of their differences with traditional frameworks. We conducted an online survey with developers of the game engines to understand why they built such engines and their opinions about the differences (if any) between engines and frameworks. 

Question 1 contains a predefined set of answers that we compiled from the literature and from the documentation and ``readme'' files studied during the manual filtering of the datasets. The respondent could choose one or more answers. We also provided a free-form text area for developers to provide a different answer and--or explain their answers. 
With Question 2, we want to understand whether game engine developers are also traditional software developers. 
Finally, Question 3 collected the developers' point of views regarding the differences (or lack thereof) between the development of engines and frameworks.


\begin{itemize}
    \item \rqEa
    \begin{itemize}
        \item Survey question 1: \mEa
        \begin{itemize}
            \item To help me to create a game
            \item To learn how to build an engine
            \item To have the full control of the environment
            \item Because the existent engines do not provide the features I need
            \item Because I wanted to work with this specific programming language
            \item Because the licenses of the existent engines are too expensive
            \item Other [please specify]
        \end{itemize}
    \end{itemize}
    \item \rqEb
    \begin{itemize}
        \item Survey question 2: \mEb [Yes or No]
    \end{itemize}
    \item \rqEc
    \begin{itemize}
        \item Survey question 3:\mEc -- [1 (very different) to 5 (very similar)]
    \end{itemize}
\end{itemize}


\section{Results from \rqA}
\label{sec:rq1-literature}


We study game engines along the literature perspective by querying both Scopus and the Internet. We report that only few works on game engines exist: mostly books, few academic papers. We did not perform a systematic literature review (SLR) \cite{Kitchenham2012} because of the small size of the current academic literature on the topic, as shown in the following.

\subsection*{\rqAa} \label{sec:rq1-literature-framework}

GitHub uses a set of ``topics''\footnote{\url{https://github.com/topics/framework}} to classify projects. It defines the topic ``framework'' as \emph{``a reusable set of libraries or classes in software. In an effort to help developers focus their work on higher level tasks, a framework provides a functional solution for lower level elements of coding. While a framework might add more code than is necessary, they also provide a reusable pattern to speed up development.''}

\citet{Pree1994} defined frameworks as having \emph{frozen} and \emph{hot} spots: code blocks that remain unchanged and others that receive user code to build the product. \citet{Larman2012} observed that frameworks use the \emph{Hollywood Principle}, ``Don't call us, we'll call you.'': user code is called by the framework. \citet{Taylor2018} sees a framework as a programmatic bridge between concepts (such as ``window'' or ``image'') and lower-level implementations. Frameworks can map architectural styles into implementation and--or provide a foundation for an architecture.

\subsection*{\rqAb} \label{sec:rq1-literature-engine}

ID Software\footnote{\url{https://www.idsoftware.com}} introduced the concept of video-game engine in 1993 to refer to the technology ``behind the game'' when they announced the game DOOM \cite{Gregory2014, Lowood2016}.
In fact, they invented the game engine around 1991 and revealed the concept around the DOOM press release early 1993 \cite{Lowood2014}.

The invention of this game technology was a discrete historical event in the early 1990s but it established MS-DOS 3.3 as a relevant gaming platform, mostly because of the NES-like horizontal scrolling emulation, allowing developers to create games similar to the ones on Nintendo console. It also introduced the separation of game engine from ``assets'' accessible to players and thereby revealed a new paradigm for game design on the PC platform \cite{Lowood2014}, allowing players to modify their games and create new experiences. This concept has since evolved into the ``fundamental software components of a computer game'', comprising its core functions, e.g., graphics rendering, audio, physics, AI \cite{Lowood2016}.

John Carmack\footnote{John Carmack was the lead programmer and co-founder of id Software in 1991.}, and to a less degree John Romero\footnote{John Romero was the designer, programmer and also co-founder of id Software in 1991.}, are credited for the creation and adoption of the term game engine. In the early 90s, they created the first game engine to \emph{separate the concerns} between the game code and its assets and to \emph{work collaboratively} on the game as a team \cite{Lowood2016, Kushner2003}. Also, they ``lent'' their engines to other game companies to allow other developers to focus only on game design.

In 2002, \citet{Lewis2002} defined game engines as ``\emph{collection[s] of modules of simulation code that do not directly specify the game's behavior (game logic) or game's environment (level data)}''. In 2007, \citet{Sherrod2007} defined engines as frameworks comprised of different tools, utilities, and interfaces that hide the low-level details of the implementations of games. Engines are extensible software that can be used as the foundations for many different games without major changes \cite{Gregory2014} and are \emph{``software frameworks for game development''}. They relieve developers so that they can focus on other aspects of game development''\footnote{\url{https://github.com/topics/game-engine}}. In 2019, \citet{Toftedahl2019} analysed and divided engines in four complementary types: (a) \emph{Core Game Engine}, (b) \emph{Game Engine}, (c) \emph{General Purpose Game Engine}, and (d) \emph{Special Purpose Game Engine}.

\subsection*{\rqAc} \label{sec:rq1-literature-related}

There are few academic papers on game engines. Most recently and most complete, \citet{Toftedahl2019} analysed the engines of games on the Steam and Itch.io platforms to create a taxonomy of game engines. They highlighted the lack of information regarding the engines used in mainstream games with only 13\% of all games reporting information about their engines. On Steam, they reported Unreal (25.6\%), Unity (13.2\%), and Source (4\%) as the main engines. On Itch.io, they observed that Unity alone has 47.3\% of adoption among independent developers.

\citet{Messaoudi2016} investigated the performance of the Unity engine in depth and reported issues with CPU and GPU consumption and modules related to rendering.

\citet{Cowan2014} in 2014 and 2016 \citep{Cowan2017}  analysed the game engines used for the development of serious games. They identified few academic sources about tools used to develop serious games. They showed that ``Second Life''\footnote{Second Life is not a game engine per se but a game that can be extended by adding new ``things'' through ``mod'' or ``modding''.} is the most mentioned game engine for serious games, followed by Unity and Unreal. They considered game engines as parts of larger infrastructures, which they call frameworks and which contain scripting modules, assets, level editors as well as the engines responsible for sound, graphics, physics, and networking. They ranked Unity, Flash, Second Life, Unreal, and XNA as the most used engines.

\citet{Neto2015} conducted a systematic literature review of game engines in the context of immersive applications for multi-projection systems, aiming at proposing a generic game engine for this purpose.

\citet{Wang2015} assumed that game development is different from traditional software development and investigated how architecture influences the creative process. They reported that the game genre significantly influences the choice of an engine. They also showed that game-engine development is driven by the creative team, which request features to the development team until the game is completed. They observed that adding scripting capability ease game-engine development through testing and prototyping.

\citet{Anderson2008} raised issues and questions regarding game engines, among which the need for a unified language of game development, the identification of software components within games, the definition of clear boundaries between game engines and games, the links between game genres and game engines, the creation of best practices for the development of game engines.


\begin{Results}{\rqA}{rqA}
We could not find many academic paper on game engines or a reliable source for gray literature. We recommend to extend this work with multivocal literature review (with academic and grey literature).
\end{Results}

\section{Results from \rqBCD}
\label{sec:study-design}

This section details the method for gathering the data and the metrics used to answer the set of RQs 2.1, 2.2, and 2.3. It also introduced the applied statistical techniques. For the sake of clarity, this section does not provide all the details but summarises the answers to each set of RQs. The Appendixes \ref{sec:rq2-static}, \ref{sec:rq3-historic}, and \ref{sec:rq4-community} present the detailed results of each set of RQs.

\subsection{Method}
\label{sec:method}

\autoref{fig:mining-method} shows the steps that we followed to mine the data to answer our questions. In Step 1, on August 8, 2019, we gathered the top 1,000 projects in GitHub related to the \emph{game-engine} and \emph{framework} topics, separately, storing each one in a specific dataset. 

In Step 2, we filtered these projects using the following criteria to remove ``noise'' and improve the quality of our dataset, which is a common approach when dealing with Github repositories \cite{Lima2014, Avelino2016} to obtain 458 engines and 743 frameworks:

\begin{figure*}[width=1\textwidth,pos=htb!]
\centering
\includegraphics[width=1\linewidth]{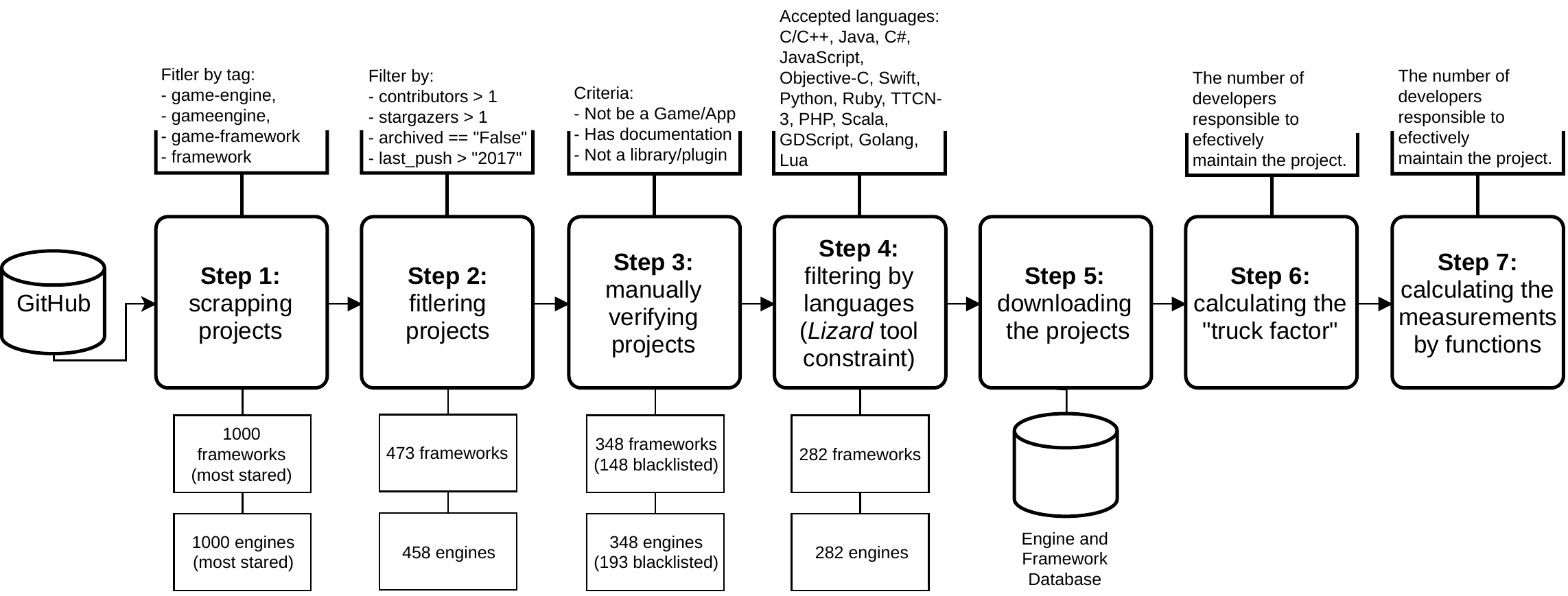}
\caption{The process used to gather the data from GitHub projects.}
\label{fig:mining-method}
\end{figure*}

\begin{itemize}
\item The project must have \emph{more than one contributor};
\item The project must have been \emph{starred at least twice};
\item The \emph{last commit} must be at least from \emph{2017};
\item The project \emph{cannot be archived}.
\end{itemize}

In Step 3, we manually analysed the remaining $458 + 743 = 1,201$ projects to remove those that are neither game engines nor frameworks according to the definitions in Section~\ref{sec:rq1-literature}. We kept 358 game engines and 358 frameworks.

In Step 4, we kept only projects with programming language supported by Lizard: C/C++, C\#, GDScript, Golang, Java, JavaScript, Lua, Objec\-tive-C, PHP, Python, Ruby, Scala, Swift, TTCN-3. We had now 282 engines and 282 frameworks.

In Step 5, we computed the metrics and stored their values in the datasets, which we describe in details in the following Section \ref{sec:metrics}.

In Step 6, we computed the \emph{truck-factor} of each project, which is the number of contributors that must quit before a project is in serious trouble \cite{Williams2002, Avelino2016}.

In Step 7, we used Lizard to gather the average value of the metrics related to functions. Lastly, we ordered the projects by popularity: how many ``stars'' they have. 

\begin{figure*}[width=1\textwidth, pos=htb!]
\centering
\includegraphics[width=.9\linewidth]{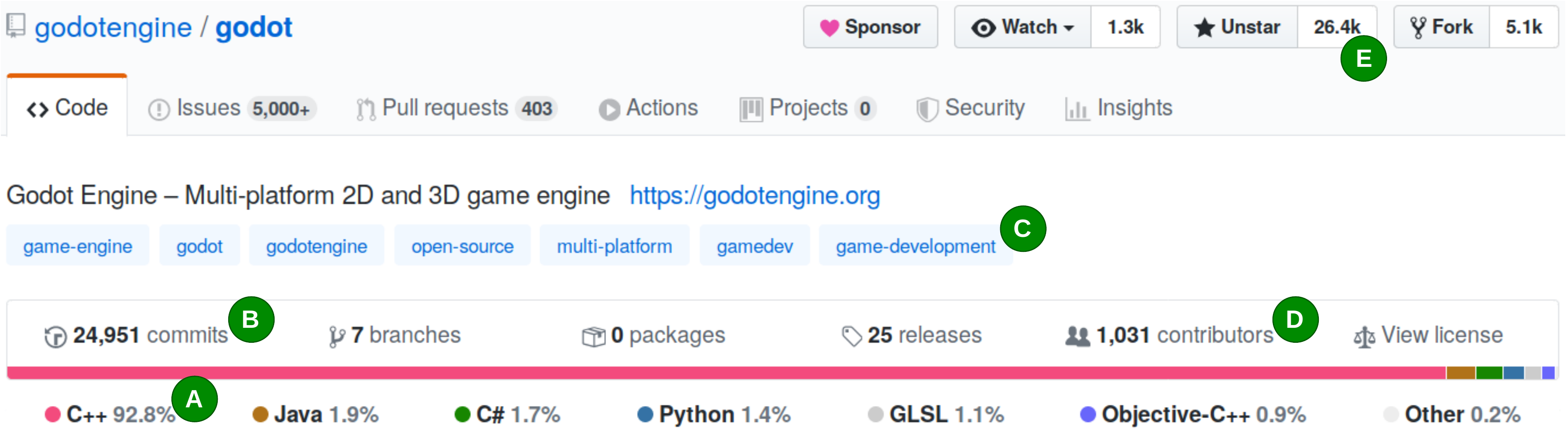}
\caption{Godot engine Github page example. In this case we considered C++ as \emph{main language} (A) and filtered the projects with least one commit from 2017 or newer (B) and with more than two contributors (D). \emph{Tags}, provided by the developers, were used to search for the engines (C). Finally, we ordered the projects by the \emph{stars} (E).}
\label{fig:godot-example}
\end{figure*}

\autoref{fig:godot-example} shows an example containing the Github page of the engine Godot\footnote{\url{https://github.com/godotengine/godot}}. We only consider the main language of the projects but most projects are composed of multiple languages. Almost all the code of Godot is written in C++ (93\%). Godot is tagged with the ``game-engine'' topic and, therefore, was found through our search. Godot is the most popular engine containing more than 26K votes.

The dataset, scripts and all the material from this study are in its replication package\footnote{\label{Footnote: Replication Package}\url{\projectpage}.}.

\subsection{Metrics}
\label{sec:metrics}

Defining metrics is challenging. Some authors warn about problems with simplistic measurements \cite{Kaner2004} and lack of precision of tools that make the measurements \cite{Lincke2008}. However, imperfect quantification is better than none \cite{DeMarco1986}.

\citet{Kaner2004} recommends the use of direct metrics\footnote{Direct metric is a metric that does not depend upon a measure of any other attribute.} but also defines a framework to described and justify the metrics.
We used a simplified version of this framework with six questions: 
\textit{Purpose} (What is the purpose of this metric?), 
\textit{Scope} (What is the scope of this metric?), 
\textit{Scale} (What is the natural scale of the attribute we are trying to metric?), 
\textit{Domain} (What is the domain this metric belongs to?), and 
\textit{Instrument} (How the metric will be measured?).

\begin{table*}[width=1\textwidth, pos=htb!]
\centering
\caption{Description of the Metrics for the Code Perspective (Rqs 2, 3, and 4). We adapted the framework defined by \citet{Kaner2004}.}
\label{tab:metrics}
\begin{tabular}{@{}%
    p{\dimexpr.05\textwidth}
    p{\dimexpr.20\textwidth}
    p{\dimexpr.08\textwidth}
    p{\dimexpr.16\textwidth}
    p{\dimexpr.06\textwidth}
    p{\dimexpr.19\textwidth}
    p{\dimexpr.12\textwidth}@{}}
\toprule
RQs      & Purpose                                    & Scope        & Metric                & Scale & Domain                        & Instrument              \\ \midrule
RQ2.1.1 &
  Verify what is the most common main languages by the projects &
  Project &
  main\_language &
  Nominal &
  Programming languages &
  API GraphQL (V4) \\ \midrule
RQ2.1.2 &
  Verify what is the most common licenses used by the projects &
  Project &
  license &
  Nominal &
  Source code licenses &
  API GraphQL (V4) \\ \midrule
\multirow[t]{3}{=}{RQ2.1.3} &
  \multirow[t]{3}{=}{Verify the project size} &
  \multirow[t]{3}{=}{Project} &
  main\_language\_size &
  Ratio &
  Positive rational numbers (Q) &
  \multirow[t]{3}{=}{API GraphQL (V4)} \\ \cmidrule(lr){4-6}
      &                                            &              & total\_size           & Ratio & Positive rational numbers (Q) &                         \\ \cmidrule(lr){4-6}
      &                                            &              & n\_file               & Ratio & Natural numbers (N)           &                         \\ \midrule
\multirow[t]{3}{=}{RQ2.1.4} &
  \multirow[t]{3}{=}{Verify the function size} &
  \multirow[t]{3}{=}{Function} &
  n\_func &
  Ratio &
  Natural numbers (N) &
  \multirow[t]{3}{=}{Lizard} \\ \cmidrule(lr){4-6}
      &                                            &              & nloc\_mean            & Ratio & Positive rational numbers (Q) &                         \\ \cmidrule(lr){4-6}
      &                                            &              & func\_per\_file\_mean & Ratio & Positive rational numbers (Q) &                         \\ \midrule
RQ2.1.5 & Verify the complexity of the function      & Function     & cc\_mean              & Ratio & Natural numbers (N)           & Lizard                  \\ \midrule

RQ2.2.1 & Verify the release strategy of the project & Project      & tags\_releases\_count & Ratio & Natural numbers (N)           & API GraphQL (V4)        \\ \midrule
RQ2.2.2 & Verify the lifetime of the project         & Project      & lifespan              & Ratio & Natural numbers (N)           & API GraphQL (V4)        \\ \midrule
\multirow[t]{2}{=}{RQ2.2.3} &
  \multirow[t]{2}{=}{Verify the contributions to the project} &
  \multirow[t]{2}{=}{Commits} &
  commits\_count &
  Ratio &
  Natural numbers (N) &
  \multirow[t]{2}{=}{Pydriller} \\ \cmidrule(lr){4-6}
      &                                            &              & commits\_per\_time    & Ratio & Positive rational numbers (Q) &                         \\ \midrule
\multirow[t]{3}{=}{RQ2.2.4} &
  \multirow[t]{3}{=}{Very the effort made in the projects} &
  \multirow[t]{3}{=}{Commits} &
  lines\_added &
  Ratio &
  Natural numbers (N) &
  \multirow[t]{3}{=}{Pydriller} \\ \cmidrule(lr){4-6}
      &                                            &              & lines\_removed        & Ratio & Natural numbers (N)           &                         \\ \cmidrule(lr){4-6}
      &                                            &              & code\_churn           & Ratio & Natural numbers (N)           &                         \\ \midrule
RQ2.3.1 & Verify the contribution of the project     & Contributors & truck\_factor         & Ratio & Natural numbers (N)           & Library Avelino et. al. \\ \midrule
\multirow[t]{2}{=}{RQ2.3.2} &
  \multirow[t]{2}{=}{Verify the popularity of the project} &
  \multirow[t]{2}{=}{Project} &
  stargazers\_count &
  Ratio &
  Natural numbers (N) &
  \multirow[t]{2}{=}{API GraphQL (V4)} \\ \cmidrule(lr){4-6}
      &                                            &              & contributors\_count   & Ratio & Natural numbers (N)           &                         \\ \midrule
\multirow[t]{3}{=}{RQ2.3.3} &
  \multirow[t]{3}{=}{Verify the how developers deal with issues in the project} &
  \multirow[t]{3}{=}{Issues} &
  issues\_count &
  Ratio &
  Natural numbers (N) &
  \multirow[t]{3}{=}{API GraphQL (V4)} \\ \cmidrule(lr){4-6}
      &                                            &              & closed\_issues\_count & Ratio & Natural numbers (N)           &                         \\ \cmidrule(lr){4-6}
      &                                            &              & closed\_issues\_rate  & Ratio & Positive rational numbers (Q) &                         \\ \bottomrule
\end{tabular}
\end{table*}

Therefore, to answer the questions \rqB, \rqC, and \rqD, we use the set of metrics described in \autoref{tab:metrics}. 
Also, the following list some details about some of the metrics.

\paragraph{RQ2.1.1. Metric: \texttt{main\_language} --}
According to Tiobe index, currently, the most common languages are C, Java, Python\footnote{\url{https://www.tiobe.com/tiobe-index/}}.
Also, GitHub uses Linguistic to determine the most common language in the project\footnote{\url{https://github.com/github/linguist}}.

\paragraph{RQ2.1.2. Metric: \texttt{license} --}
MIT, Apache-2.0, and GPL-V3 were the most common open source licenses in 2019\footnote{\url{https://bit.ly/3f4myu3}}.	
GitHub has a ``license picker'' allowing the user to choose from different open-source licenses\footnote{\url{https://bit.ly/32YNOYv}}.

\paragraph{RQ2.1.3. Metrics: \texttt{main\_language\_size, total\_size, n\_file} --}
GitHub recommends repositories with less than 1GB, and less than 5GB is strongly recommended. Also with 100MB maximum file size limit\footnote{\url{https://bit.ly/2BDuKns}}.

\paragraph{RQ2.1.4. Metrics: \texttt{n\_func, nloc\_mean, func\_per\_file\_mean} -- }
Lizard\footnote{\url{https://github.com/terryyin/lizard}} gives a list of all functions in the project with NLOC (lines of code without comments) and CCN (cyclomatic complexity number). It also gives the list of files and the functions' name (signatures).

\paragraph{RQ2.1.5. Metrics: \texttt{cc\_mean} --}
McCabe's \cite{McCabe1976} recommends keeping the complexity of modules below 10.
 
\paragraph{RQ2.2.1. Metric \texttt{tags\_releases\_count}:} 
Within the context of GitHub, a tag is a release of the product.

\paragraph{RQ2.2.2. Metrics \texttt{commits\_count, commits\_per\_time} --}
PyDriller is a Python framework to analyze Git repositories\footnote{\url{https://github.com/ishepard/pydriller}}.

\paragraph{RQ2.2.4. Metrics: \texttt{lines\_added}, \texttt{lines\_removed}, \texttt{code\_churn} -- }
Code churn measures the number the amount of code changes occurred during development of code \cite{Shin2011}.
We use the sum of deleted and removed lines.

\paragraph{RQ2.3.1. Metric: \texttt{truck\_factor} --}
According to \citet{Avelino2016},  truck-factor is the number of people on the team that have to be hit by a truck before the project becomes problematic. Therefore, systems with low truck-factor have problems with strong dependency of certain developers.
Linux has truck-factor of 57 and Git 12.
We used a library defined by \citet{Avelino2016} as instrument.

\paragraph{RQ2.3.2. Metrics: \texttt{stargazers\_count} --} 
``vuejs'' is the most stared github opensource software project with more than 169K\footnote{\url{https://github.com/vuejs/vue}}.
In 2019, ``microsoft/vscode'' had the highest number of contributors with 19.1K\footnote{\url{https://octoverse.github.com/}}.

\subsection{Analysis} 
\label{sec:method-analysis}

We used the statistical-analysis workflow-model for empirical software-engineering research \cite{DeOliveiraNeto2019} to test statistically the differences between engines and frameworks. For each continuous variable, we used descriptive statistics in the form of tables with mean, median, min, and max values, together with boxplots. For the boxplots, to better show the distributions, we removed outliers using the standard coefficient of \emph{1.5} ($Q3 + 1.5 \times IQR$). We observed outliers for all the measures, with medians skewed towards the upper quartile (Q3). To check for normality, we applied the Shapiro test \cite{Wohlin2012} and checked visually using Q--Q plots. Normality < 0.05 means the data is not normally distributed. Finally, given the data distribution, we applied the appropriate statistical tests and computed their effect sizes.

\subsection{Results for \rqB} \label{sec-res-summary-rqb}


\begin{table}[width=1\linewidth, pos=htb!]
\caption{Statistical Tests, \rqB}
\label{tab:mw-rq2}
\centering
\begin{tabular*}{\tblwidth}{@{}Lrrr@{}}
\toprule
Variable & P-value & Estimate & Effect \\ \midrule
main\_language\_size & \textless 0.01 & 0.28 & 0.189 (small) \\
total\_size & \textless 0.01 & 0.34 & 0.188 (small) \\
n\_file & \textless 0.01 & 45.00 & 0.155 (small) \\
n\_func & \textless 0.01 & 769.00 & 0.211 (small) \\
nloc\_mean & \textless 0.01 & 2.12 & 0.297 (small) \\
func\_per\_file\_mean & \textless 0.01 & 3.13 & 0.208 (small) \\
cc\_mean & \textless 0.01 & 0.53 & 0.356 (small) \\
\bottomrule
\end{tabular*}
\end{table}

\autoref{tab:mw-rq2} shows the results of Wilcoxon tests. The p-values $< 0.01$ indicate that the distributions are not equal and there is a significant difference between engines and frameworks, although this difference is \emph{small}. The biggest  effects are related to source code metrics, i.e., \emph{nloc\_mean} and \emph{cc\_mean}.

The implementation of game engines and traditional frameworks are different but without statistical significance. Engines are bigger and more complex than frameworks. They use mostly compiled programming languages vs.\ interpreted ones for frameworks. They both often use the MIT license.

\subsection{Results for \rqC} \label{sec-res-summary-rqc}

Overall, all metrics have similar median values when comparing both groups, except for \emph{tags\_releases\_count}. In fact, engines releases way less versions (median is one) than frameworks (median is 32).

\begin{table}[width=1\linewidth,pos=htb!]
\caption{Statistical Tests, \rqC}
\label{tab:mw-rq3}
\centering
\begin{tabular*}{\tblwidth}{@{}Lrrr@{}}
\toprule
Variable & P-value & Estimate & Effect \\ \midrule
tags\_releases\_count & \textless 0.01 & -24.00 & -0.613 (large) \\
lifespan & \textless 0.01 & -56.29 & -0.32 (large) \\
commits\_count & \textless 0.01 & -175.00 & -0.198 (large) \\
commits\_per\_time & \textless 0.01 & -0.30 & -0.198 (large) \\
lines\_added   & \textless 0.01 & 134.47 & 0.219 (small) \\
lines\_removed & \textless 0.01 & 48.83  & 0.168 (small) \\ 
cchurn\_delta  & \textless 0.01 & 153.88 & 0.224 (small) \\ 
cchurn\_sum    & \textless 0.01 & 222.48 & 0.212 (small) \\
\bottomrule
\end{tabular*}
\end{table}

\autoref{tab:mw-rq3} shows the results of Wilcoxon tests, showing large differences for all historical measures except \texttt{lines\_added, lines\_removed,} and \texttt{code\_churn}. 
Versioning does not look like a well-followed practice in engine development, with few versions compared to frameworks. Commits are less frequent and less numerous in engines, which are younger and have shorter lifetimes when compared to frameworks.

\subsection{Results for \rqD} \label{sec-res-summary-rqd}


\begin{table}[width=1\linewidth, pos=htb!]
\caption{Statistical Tests, \rqD}
\label{tab:mw-rq4}
\centering
\begin{tabular*}{\tblwidth}{@{}Lrrr@{}}
\toprule
Variable & P-value & Estimate & Effect \\ \midrule
stargazers\_count & \textless 0.01 & -358.00 & -0.511 (large) \\
contributors\_count & \textless 0.01 & -9.00 & -0.459 (large) \\
truck\_factor & 0.01 & \textless 0.01 & -0.138 (large) \\
issues\_count & -139.00 & \textless 0.01 & -0.451 (large) \\
closed\_issues\_count & -122.00 & \textless 0.01 & -0.459 (large)\\
closed\_issues\_rate & -0.05 & \textless 0.01 & -0.27 (large)\\
\bottomrule
\end{tabular*}
\end{table}

\autoref{tab:mw-rq4} shows the results of Wilcoxon tests, indicating a large difference in all measures related to community. 
The truck-factor shows that the majority of the projects have few contributors. Some uncommon languages, like Go and C\#, are popular compared to others in more prevalent projects, e.g., C++ and JavaScript.

\begin{Results}{\rqBCD}{rqBCD}
We observed that, although \textit{static} characteristics of game engines and frameworks are similar, the \textit{community} of these projects differ. Also, \textit{historical} aspects are mixed, as engines have a smaller lifespan and less releases, yet similar effort on commits contribution and code churn.
\end{Results}


\section{Results from \rqE}
\label{sec:rq5-human}

We now discuss developers' own perception of game engines and of their differences with traditional frameworks.
We used an online form to contact developers over a period of three days. We sent e-mails to 400 developers of the game engines in our dataset, using the truck-factor of each project: developers who collaborate(d) most to the projects. We received 124 responses, i.e., 31\% of the developers.
The survey, answers, and scripts for their analyses are in the replication package\textsuperscript{\ref{Footnote: Replication Package}}.





\subsection*{Question 1: \mEa}

\begin{figure}[pos=htb!]
\centering
\includegraphics[width=.9\linewidth]{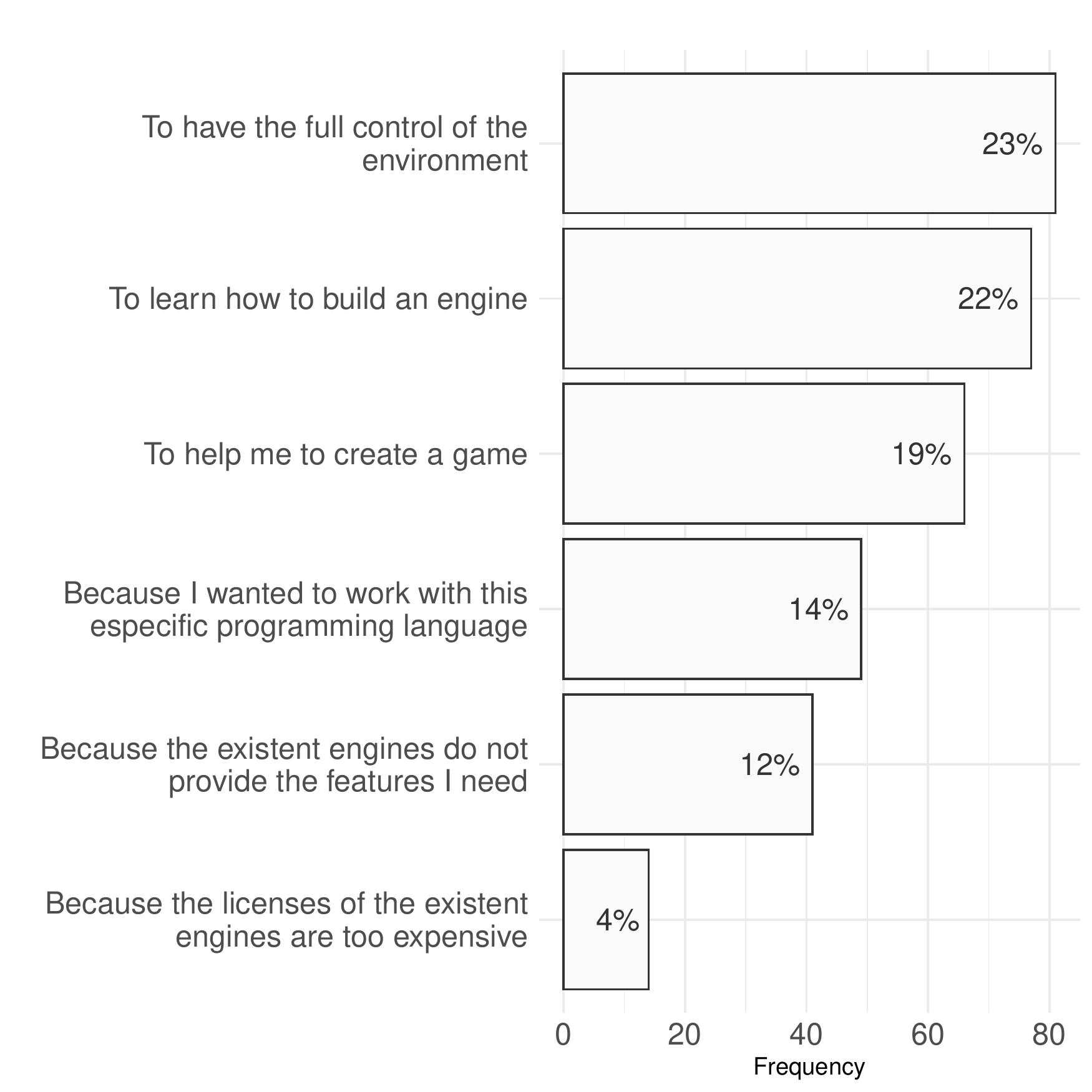}
\caption{Answers to Question 1: \mEa}
\label{fig:survey-q1}
\end{figure}

\autoref{fig:survey-q1} shows the breakdown of the developers' answers. Having access to the source code, freedom to develop, etc., i.e., \emph{control of the environment}, is the developers' major reason for working on a game engine while \emph{learning} to build an engine is the second reason; explaining why many engines have few developers and commits.

The third reason is to build a \emph{game}, confirming the lack of clear separation between developers and game designers. It is indeed common for game developers to act also as game designers, specially in independent games, e.g., the single developer of Stardew Valley\footnote{\url{https://www.stardewvalley.net/}}.

The next answer is about working with a specific \emph{language}, also related with learning: when learning a new language, developers want to apply or test their knowledge on some projects, and game engines are interesting candidates.

Also related to the environment, the next answer concerns the \emph{features} offered by existing engines: reusing or creating a new engine may be necessary for certain, particular games with specific requirements. Developers think as game designers: the game concept(s) may require a new engine.

The engine \emph{licenses} are the least concern: fees and taxes from vendors, e.g., Unreal and Unity, are not important to developers because some licenses are ``indie'' friendly and offer low rates for indie games \citep{Toftedahl2019}.

Finally, 19 developers provided ``Other'' answers: they work on game engines because ``it is fun'' and--or they have access to some source code, e.g., one developer who reverse-engineered a proprietary engine wrote:

\begin{quote}
\textit{``The source for the original engine was proprietary and so we opened the platform by reverse-engineering it then re-implementing under GPL3.''}
\end{quote}

Other answers include performance, platform compatibility, new experimental features, and creating a portfolio.

\subsection*{Question 2: \mEb}

The great majority of developers, 119 of the 124 respondents (96\%), have experience with traditional software. The respondents can be considered general software developers with expertise in engine development.

\subsection*{Question 3: \mEc}

\begin{figure}[pos=htb!]
\includegraphics[width=1\linewidth]{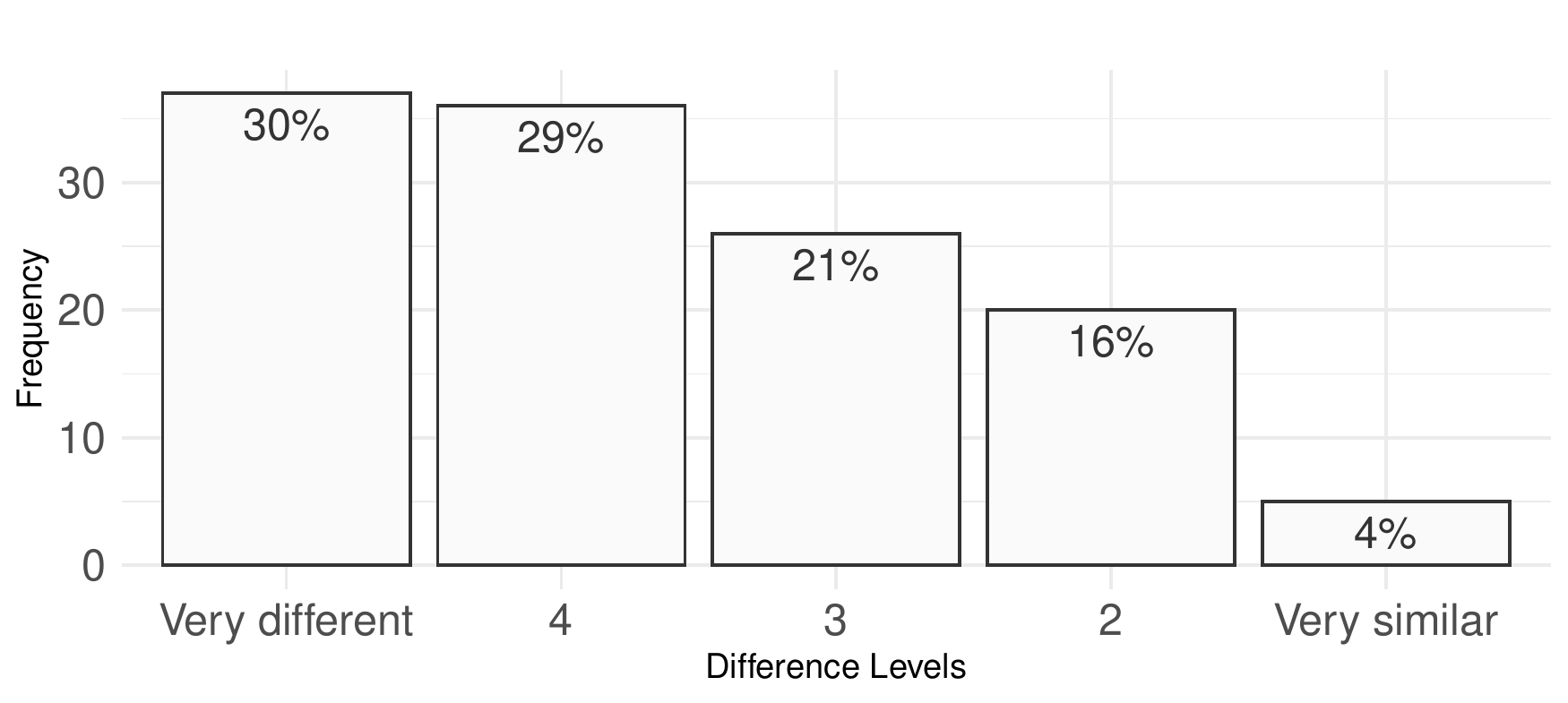}
\caption{Answers to Question 3: \mEc}
\label{fig:survey-q3}
\end{figure}

\autoref{fig:survey-q3} shows that engine developers consider engines different from frameworks: 59\% of the respondents believe that engines follows a different process from frameworks. Only 20\% believe this this process is similar. This is a surprising result as they also have experience in developing traditional software.

\begin{Results}{\rqE}{rqE}
The developers' main reasons to work on an engine is (a) having better control over the environment and source code, (b) learning game-engine development, and (c) helping develop a specific game. Almost all the engine developers have experience with traditional software. They consider these two types of software as different.
\end{Results}


\section{Discussions}
\label{sec:disc}

We now discuss the results of our study of engines along the three perspectives.

\subsection{Perils for Engines and Frameworks}

\citet{Kalliamvakou2014} analysed developers' usages of GitHub and reported a set of perils and promises related to GitHub projects. \autoref{tab:perils} shows the perils applying to the objects of our study: engines and frameworks. The perils 7, 8 (about pull-requests), and 9 (activity outside GitHub) are out of scope of our dataset.

\begin{table}[pos=htb!]
\caption{Perils of Github repositories adapted from \citet{Kalliamvakou2014}. Perils 7, 8, and 9 do not pertain to this work.}
\label{tab:perils}
\centering
\begin{tabular}{@{}
p{\dimexpr.01\textwidth}
p{\dimexpr.3\textwidth}
p{\dimexpr.05\textwidth}
p{\dimexpr.05\textwidth}
@{}}
\toprule
\# & Perils & Eng. & Fram. \\ \midrule
1 & A repository is not necessarily a project. & False & False \\
2 & Most projects have very few commits. & False & False \\
3 & Most projects are inactive. & False & False \\
4 & A large portion of repositories are not for software development. & True & False \\
5 & Two thirds of projects (71.6\% of repositories) are personal. & True & False \\
6 & Only a fraction of projects use pull requests. And of those that use them, their use is very skewed. & True & True \\ \bottomrule
\end{tabular}
\end{table}

In Peril 1, the authors distinguished forks and base repositories. In our search, we observed that most repositories are base ones. We found few forks that we removed during the manual filtering. Therefore, this peril is false for both engines and frameworks.

In Peril 2, the authors reported that the median of commits were 6, with 90\% of the projects having less than 50 commits. We observed that the engines and frameworks in our dataset have medians of 616 and 833 commits, respectively, as shown in \autoref{tab:ds-rq3}.

Peril 3 does not apply to our dataset as the projects have a median of one commit per week and because we removed projects with more than two years without commits.

For Peril 4, the authors found that about 10\% of the developers used GitHub for storage. This is partially true for our dataset: we found engine repositories that were used mostly to store assets, documentation, and other files.

Peril 5 is present in this study, specifically in game engine projects: although we removed engines with less than 2 contributors, we found many warnings in read-me files stating that an engine was only for ``personal use'', an ``unfinished project'', or for ``educational purposes'' only.

Peril 6 is true for all projects. The number of pull requests for engines is lower than that for frameworks: at least 50\% of the engine projects have at least 10 closed pull requests, while frameworks have 100s.

In general, the perils found in any repositories in GitHub do not apply to our dataset. Engines and frameworks seem different to the projects studied by \citet{Kalliamvakou2014}.

\subsection{Discussion of \rqA}

In theory, game engines and frameworks have similar objectives: they are \emph{modular platforms} for \emph{reuse} that provide a \emph{standard} way to develop a product, lowering the barrier of entry for developers by \emph{abstracting} implementation details.

We could classify frameworks in different categories, according to their domains, e.g., Web apps, mobile apps, AI, etc. In a same category and across categories, two frameworks are not the same. They provide their functionalities in different ways. Similarly, game engines also belong to different categories and are different from one another. For example, 3D or 2D and specific for game genres, like \textit{platformer}, \textit{shooter}, \textit{racing}, etc.

Traditional frameworks provide business services while game engines support entertaining games \cite{Kasurinen2017}. The process of finding the ``fun factor'' is exclusive to game development \cite{Lewis2011, Callele2015} but do not exempt developers from using traditional software-engineering practices \cite{Petrillo2009, Kasurinen2016}. Game engines are tools that help game developers to build games and, therefore, are not directly concerned with non-functional requirements of games, such as ``being fun''.

\begin{figure}[pos=htb!]
\centering
\includegraphics[width=.8\linewidth]{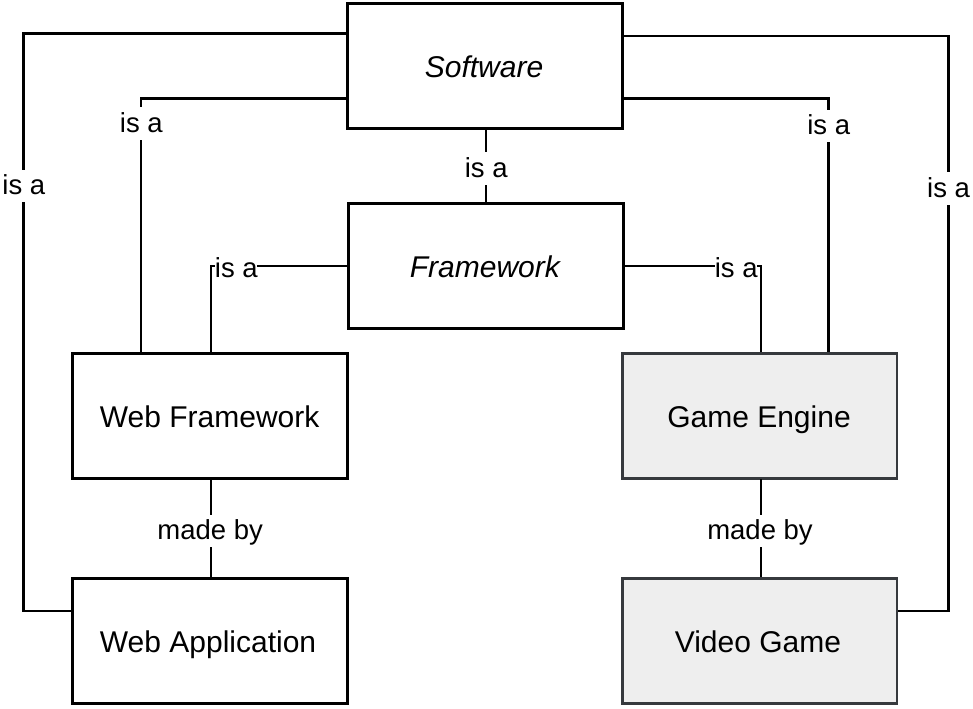}
\caption{Semantic relationship between software, framework, and product.}
\label{fig:taxonomy-engine}
\end{figure}

\autoref{fig:taxonomy-engine} shows the relationship between game engines, games, frameworks, and traditional software: a \emph{video game} is a product built on top of a \emph{engine}, like a \emph{Web app} is built on top of a \emph{Web framework}. Engines are a specific kind of framework used to build games. Everything described is a \emph{software}: Scrumpy\footnote{\url{https://scrumpy.io}} is a Web app written with Vue while Dota 2\footnote{\url{http://blog.dota2.com}} is a game made with Source.

\subsection{Discussion of \rqBCD}

\subsubsection{\rqB}

\paragraph{Differences in Programming Languages.} There is a discrepancy between the languages used in game engines, which belong mostly to the C family, and frameworks, developed mostly with interpreted languages. We explain this difference as follows: engines must work close to the hardware and manage memory for performance. Low-level, compiled languages allow developers to control fully the  hardware and memory. Frameworks use languages providing higher-level abstractions, allowing developers to focus on features. Frameworks and engines are tools on which developers build their products, who choose the most effective language for their needs.

This observation highlights the needs for performance in engines, through low-level communication with hardware and memory. With the rise of WebAssembly\footnote{\url{https://webassembly.org/}} and the possibility of running compiled code in Web browsers, this observation could change in the near future.

We explain the predominance of C++ for engines by a set of features of this language: abstraction, performance, memory management, platforms support, existing libraries, and community. These features together make C++ a good choice for game developers.

Engines are usually written (or extended) via their main programming language. However, to ease the design, implement, test workflow during production, game developers often add scripting capabilities to their engines. Therefore, when writing a game, game developers may not code directly with low-level languages but use scripts; sometimes with in a specific domain-specific language. 
For example, Unity, although written in C++, offers scripting capabilities in C\#\footnote{\url{https://docs.unity3d.com/Manual/ScriptingSection.html}} for game developers to build their games. Furthermore, the developer can, possible, finish its game just using this high level language. For any further extension in the game engine they will need to deal with the low level language. On the other hand, frameworks rarely offer scripting capabilities: their products are often written in the same programming languages.

\paragraph{Similarities in Licenses.}

The MIT License is the most used license by both open-source frameworks and open-source engines because it allows reusing and distributing source and--or compiled code in open or proprietary products, with or without changes. Developers can use such these frameworks and engines to create and distribute their software and games without restriction. Also, they can extend or change the code without having to share their intellectual property.

\paragraph{Similarities in Sizes and Complexities.}

Our results show small differences in sizes and complexities between engines and frameworks, yet not enough to consider engines different from frameworks. 

The size of a piece of code is a simplistic proxy to its quality. 
Also, regarding the languages and numbers of files, we expected that larger values for frameworks, given the numbers of configuration files and testing functions. However, we reported that engines are larger in all cases, although by a small margin.

The complexities of the functions was another surprise given the large number of small engines: engines are more complex, although by a small difference.

\subsubsection{\rqC}

Our results showed that 40\% of the engines do not have tags, which could mean that they are still under development and no build is available. 

However, our dataset contains the most important game engines on GitHub, thus there should be other reasons for the lack of engine releases. During our manual analysis, we found engines with warning messages alerting that they were incomplete, lacking some essential features. Also, we observed that about one third of the engines have only two collaborators. This fact combined with the complexity of engines could explain the difficulty to release a first feature-complete version.

Frameworks are released more often than engines with more commits performed more regularly. There are thus meaningful differences between engines and frameworks, which could be explained by the higher popularity of the frameworks (see next section).

\subsubsection{\rqD}

\paragraph{Differences in Truck Factor.}

The truck-factor is 1 for most of the engines (83\%). \citet{Lavallee2015} considered that, in addition to being a threat to a project survival, a low truck-factor causes also delays, as the knowledge is concentrated in one developer only. This concentration further limits adoption by new developers. We believe that low truck-factor values are due to the nature of the engines, i.e., side/hobby projects. In contrast, popular frameworks do not have such a dependency on single developers.

\paragraph{Differences in Community Engagement.}

We assumed that the numbers of stars for projects in GitHub are a good proxy for their popularity \cite{Borges2018}. Surprisingly, engines written in Go and frameworks written in C\# are most popular, even though their total numbers are low. JavaScript and C are second and third, respectively. Java is barely present despite its age and general popularity.

\subsection{Closed-Source vs.\ Open-Source Game Engines}

The great majority of commercial games are written with proprietary, closed-source game engines. Recently, an effort for building a robust and powerful open-source game engine became popular with the Godot project. The Godot game engine was thus used by many indie games\footnote{\url{https://itch.io/games/made-with-godot}}, some of them of high quality\footnote{\url{https://youtu.be/UEDEIksGEjQ}}. 

Open-source tools for game development had a promising start with Doom and its engine. However, the game industry took another route and closed-source engines are commonplace nowadays. Despite the difference in popularity between open-source game engines and traditional, open-source frameworks, we believe that open-source is the right path to follow. It democratizes and allow a soft learning curve for beginners. Also, it will allow the creation of more diverse games.

\subsection{Threats to Validity}

Our results are subject to threat to their internal, construct, and external validity.

\paragraph{Internal Validity.} We related engines and frameworks with static and historical measures. As previous works, we assumed that these measures represent the characteristics that they measure as perceived by developers. It is possible that other measures would be more relevant and representative for developers' choices and perceptions. We mitigated this threat by exploring different perspectives: \emph{literature}, \emph{code}, and \emph{human}. Also, we divided measures along different aspects (static, historical, and community).

\paragraph{Construct Validity.} We assumed that we could compare fairly projects in different programming languages, for different domains, and with different purposes, as in various previous studies. We claim that different projects can be compared by considering these projects from three different perspectives: \emph{literature}, \emph{code}, and \emph{human}.

\paragraph{External Validity.} We studied only open-source projects accessible to other researchers and to provide uniquely identifying information. We also shared on-line\textsuperscript{\ref{Footnote: Replication Package}} all the collected data to mitigate this threat by allowing others to study, reproduce, and complement our results.

\paragraph{Conclusion Validity.} We did not perform a systematic literature review integrating gray literature available on the Internet. We accept this threat and plan a multivocal literature review in future work. Our study of the literature confirmed that game engines are little studied in academia.

The higher popularity of the frameworks is a concern: the numbers of contributors are larger and could lead to unfair comparisons. We ordered the dataset by the most popular frameworks and engines, so we expected such effect. In the future, we will improve the categorization of our dataset by separating frameworks and engines based on their domains (Web, security, etc., and 2D and 3D games, etc.).

We mined the dataset using the tags of GitHub with which developers classify their projects. For game engines, we used some variations like \emph{game engine}, \emph{game-engine}, or \emph{ga\-me\-en\-gine}. We may have missed some projects if developers did not use relevant, recognisable tags. For example, the game engine Piston\footnote{\url{https://github.com/PistonDevelopers/piston}}, written in Rust by 67 contributors, is not part of our dataset because it was tagged as \emph{``piston, rust, modular-game-engine''}. However, we claim that such engines are rare and their absence does not affect our results based on 282 engines and 282 frameworks.

Regarding our survey, Question 3 is broad and could have mislead developers. Although the requirements are different, developers are still creating the building blocks that will serve to build a product. We mitigated this threat through Questions 1 and 2 and the other two perspectives.

Even after filtering out projects with at less than two contributors, most of the open-source engines are, in fact, personal projects. A few, popular, open-source game engines are used by the majority of the released games. In general, commercial games are built using proprietary/closed-source engines (Unity and Unreal). In future work, we could use other metrics to filter the projects than their numbers of contributors and stars; for example, the numbers and stars of the games released using these engines.

Our conclusions, in particular those stemming from RQ2 and RQ3, depend on the \emph{users} of the analysed projects: a project with a larger user base would certainly receive more bug reports, see more commits and more contributions, etc. Therefore, comparing the top frameworks, which are certainly used in dozens of others projects, and the top game engines, which are mostly ``personal'' projects, could be unfair to game engines. We must accept this threat in the absence of means to obtain statistics on the user base of GitHub projects. Future work includes possibly using GitHub Insights on a selected set of projects in collaboration with their developers' teams.

Finally, we acknowledge that a great part of game-engine development is closed-source. Therefore, these results might not generalize to game engines overall but should hold true to open-source game engines.


\section{Conclusion}
\label{sec:conclusion}

This paper is a step towards confirming that software-engineering practices apply to game development given their commonalities. It investigated open-source game engines, which form the foundation of video games, and compared them with traditional open-source frameworks. Frameworks are used by developers to ease software development and to focus on their products rather than on implementation details. Similarly, game engines help developers create video games and avoid duplication of code and effort. 

We studied open-source game engines along three perspectives: \emph{literature}, \emph{code}, and \emph{human}. Our literature review showed a lack of academic studies about engines, especially their characteristics and architectures. 
Yet, we showed that, different from what researchers and engine developers think, there are qualitative but no quantitative differences between open-source engines and open-source frameworks. Hence, game engines must be an object of study in software-engineering research in their own right.

We divided the code perspective into three points of view: static code (RQ2.1), history of the projects (RQ2.2), and of their community characteristics (RQ2.3). We studied 282 engines and 282 frameworks from GitHub and contributed with the first corpus of curated open-source engines\textsuperscript{\ref{Footnote: Replication Package}}. We reported no significant difference between engines and frameworks for size and complexity but major differences in terms of popularity and community engagement. The programming languages adoption differed greatly also with engines mostly written in C, C++, and C\# and frameworks mostly in Java\-Script, PHP, and Python. We observed that engines have shorter histories and fewer releases than frameworks.

Finally, our survey results showed that engine developers have also experience in developing traditional software and that they believe that game engines are different from frameworks. The developers' objectives for developing engines are (a) better control the environment and source code, (b) learn, and (c) develop specific games.

\begin{summarybox}
We conclude that open-source game engines share similarities with open-source frameworks, mostly regarding their concepts, code characteristics, and contribution effort. Yet, while engines projects are mainly personal, the communities around framework projects are larger, with longer lifespans, more releases, better truck-factor, and more popularity.
\newline
\newline
Therefore, engine developers should adopt the similar software code-quality toolkit when dealing with code.
Finally, the low truck-factor and smaller user-base suggests that more care should be given to the documentation of the open-source engines.



\end{summarybox}

We will also consider contacting a subset of the top projects and work with their developers' teams so that they install GitHub Insights, which would provide further information on the projects, unavailable at the moment when analysing GitHub data as ``outsiders'' to the projects.

Some engines appears suitable for a deeper investigation of their core architectures. The outliers are good candidates to find anti-patterns related to engines and frameworks. While \citet{Gregory2014} presented a complex description of the architecture of an engine, it would be interesting to see how a real, successful engine architecture is similar to the one proposed by the author. Also, we did not discuss in details the most popular, closed-source engines: Unity and Unreal. We could also study the differences between engines and frameworks regarding their workflow to reveal new differences between both types of software. Finally, further investigate engines and frameworks communities (developers' turnover and how teams are geo-dispersed) as well as why and how these projects choose their languages.

\section*{Acknowledgement}

The authors thank all the anonymous developers for their time. The authors were partly supported by the NSERC Discovery Grant and Canada Research Chairs programs.

\clearpage

\bibliographystyle{unsrtnat}
\bibliography{lib-engine.bib}

\clearpage

\appendix

\section*{Appendix}

\begin{figure*}
\centering
\includegraphics[width=1\textwidth]{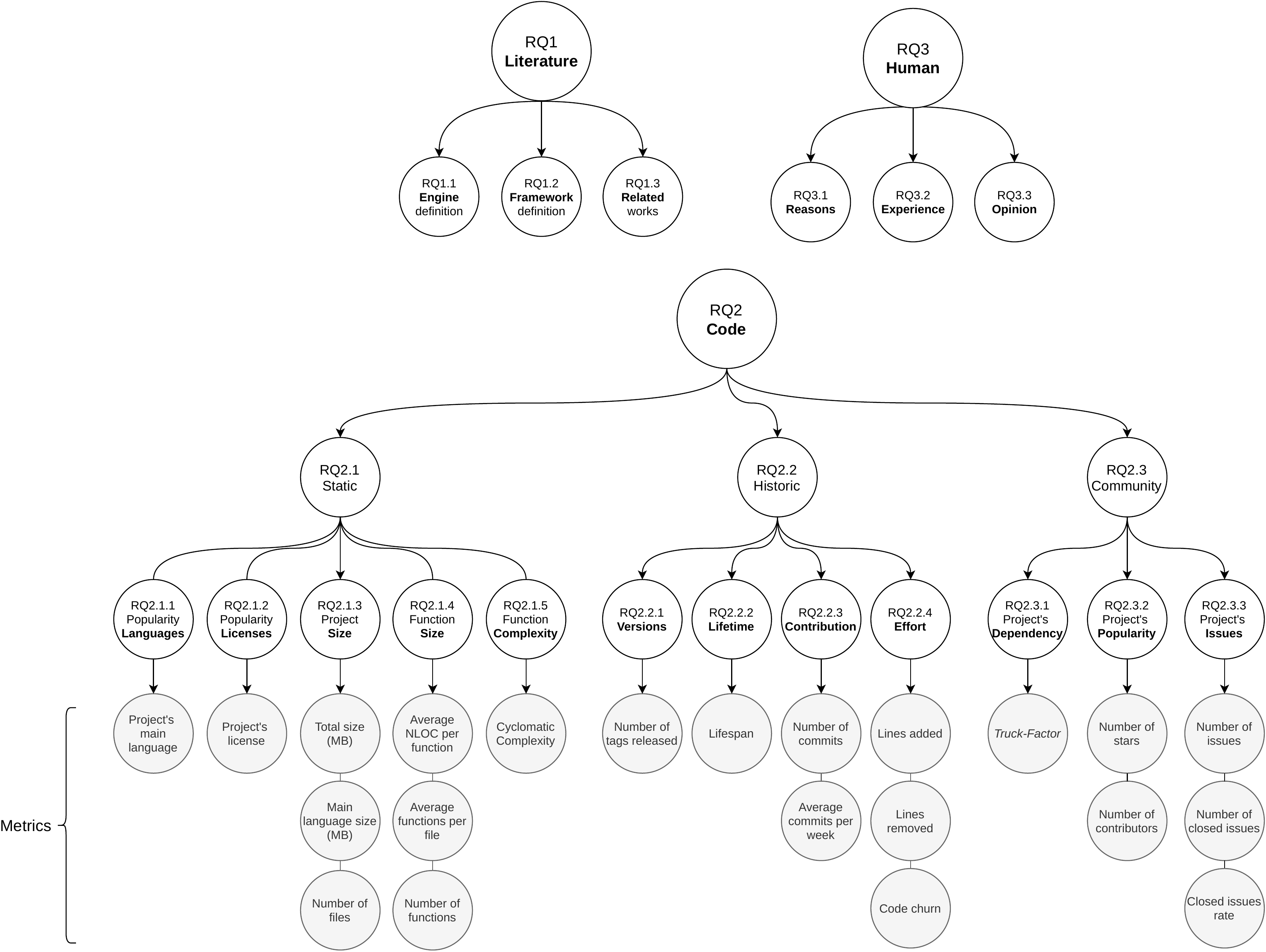}
\caption{The three perspectives of the study, the research questions and metrics.}
\label{fig:perspectives}
\end{figure*}


\section{Detailed Results for \rqB}
\label{sec:rq2-static}

\begin{table*}[width=1\textwidth, pos=tb!]
\caption{Descriptive Statistics, \rqB. Normality \textless 0.01 means the data is not normally distributed.}
\label{tab:ds-rq2}
\centering
\begin{tabular*}{\tblwidth}{@{}LLlrrrrrr@{}}
\toprule
RQs & Variable & Type & \multicolumn{1}{l}{Mean} & \multicolumn{1}{l}{Std.Dev.} & \multicolumn{1}{l}{Median} & \multicolumn{1}{l}{Min} & \multicolumn{1}{l}{Max} & \multicolumn{1}{l}{Normality} \\ \midrule

\multirow{6}{*}{RQ2.3} & main\_language\_size & engine  &  5.78 & 14.95 & 1.09 & 0.00 & 102.10 & \textless 0.01 \\
 & main\_language\_size & framework & 3.82 & 17.74 & 0.55 & 0.00 & 276.76 & \textless 0.01 \\ \addlinespace
 & total\_size & engine  & 7.66 & 20.31 & 1.22 & 0.00 & 155.32 & \textless 0.01 \\
 & total\_size & framework & 4.82 & 26.27 & 0.60 & 0.00 & 423.37 & \textless 0.01 \\ \addlinespace
 & n\_file & engine  & 685.60 & 1853.12 & 171.00 & 1.00 & 23379.00 & \textless 0.01 \\
 & n\_file & framework & 456.93 & 1053.45 & 97.50 & 1.00 & 8062.00 & \textless 0.01 \\ \midrule

\multirow{6}{*}{RQ2.4} & n\_func & engine  & 10130.74 & 21627.98 & 2394.00 & 1.00 & 163779.00 & \textless 0.01 \\
 & n\_func & framework & 5924.17 & 14469.19 & 960.50 & 1.00 & 145288.00 & \textless 0.01 \\ \addlinespace
 & nloc\_mean & engine  & 12.94 & 15.17 & 11.07 & 1.32 & 247.25 & \textless 0.01 \\
 & nloc\_mean & framework & 12.71 & 33.89 & 8.79 & 1.00 & 539.79 & \textless 0.01 \\ \addlinespace
 & func\_per\_file\_mean & engine  & 20.60 & 40.32 & 12.34 & 1.00 & 370.14 & \textless 0.01 \\
 & func\_per\_file\_mean & framework & 23.03 & 82.62 & 8.57 & 1.00 & 1070.13 & \textless 0.01 \\ \midrule

\multirow{2}{*}{RQ2.5} & cc\_mean & engine  & 3.09 & 2.35 & 2.77 & 1.00 & 36.19 & \textless 0.01 \\
 & cc\_mean & framework & 2.68 & 3.75 & 2.14 & 1.00 & 60.22 & \textless 0.01 \\
\bottomrule
\end{tabular*}
\end{table*}

\subsection*{\rqBa}

\autoref{tab:languages} shows the popularity of the programming languages in both framework and game engines, ordered by the numbers of projects. The most used languages in game engines belong to the C family: C, C++, and C\#. Together, they represent about 64\% of the code. For frameworks, Java\-Script, PHP, and Python are the most used languages with 51\% of the code. C++ and JavaScript are the most used language for games and frameworks.

\begin{table}[width=1\linewidth, pos=htb!]
\caption{Popularity of programming languages among engines and frameworks.}
\label{tab:languages}
\centering
\begin{tabular*}{\tblwidth}{@{}Lrrrrrr@{}}
\toprule
 & \multicolumn{2}{c}{Engine} & \multicolumn{2}{c}{Framework} & \multicolumn{2}{c}{Total} \\ \cmidrule(l){2-7}
 & \multicolumn{1}{c}{N} & \multicolumn{1}{c}{\%} & \multicolumn{1}{c}{N} & \multicolumn{1}{c}{\%} & \multicolumn{1}{c}{N} & \multicolumn{1}{c}{\%} \\ \midrule
C++ & 107 & 37.94\% & 10 & 3.55\% & 117 & 20.74\% \\
JavaScript & 28 & 9.93\% & 71 & 25.18\% & 99 & 17.55\% \\
Python & 14 & 4.96\% & 45 & 15.96\% & 59 & 10.46\% \\
C & 41 & 14.54\% & 11 & 3.90\% & 52 & 9.22\% \\
PHP & 3 & 1.06\% & 46 & 16.31\% & 49 & 8.69\% \\
C\# & 33 & 11.70\% & 15 & 5.32\% & 48 & 8.51\% \\
Java & 27 & 9.57\% & 19 & 6.74\% & 46 & 8.16\% \\
Go & 14 & 4.96\% & 21 & 7.45\% & 35 & 6.21\% \\
TypeScript & 7 & 2.48\% & 18 & 6.38\% & 25 & 4.43\% \\
Swift & 2 & 0.71\% & 13 & 4.61\% & 15 & 2.66\% \\
Scala & 1 & 0.35\% & 5 & 1.77\% & 6 & 1.06\% \\
Objective-C & 1 & 0.35\% & 4 & 1.42\% & 5 & 0.89\% \\
Lua & 4 & 1.42\% & 0 & 0.00\% & 4 & 0.71\% \\
Ruby & 0 & 0.00\% & 4 & 1.42\% & 4 & 0.71\% \\ \bottomrule
\end{tabular*}
\end{table}

The distributions of the languages differ for each group. Game engines are mainly built in C++. For frameworks, the differences between the top three languages are smaller. Because we sorted projects by popularity and most top frameworks focus on Web development, interpreted languages are used in the majority of their code.

\begin{figure}[pos=htb!]
\centering
\begin{subfigure}{1\linewidth}
\centering
\includegraphics[width=.9\linewidth]{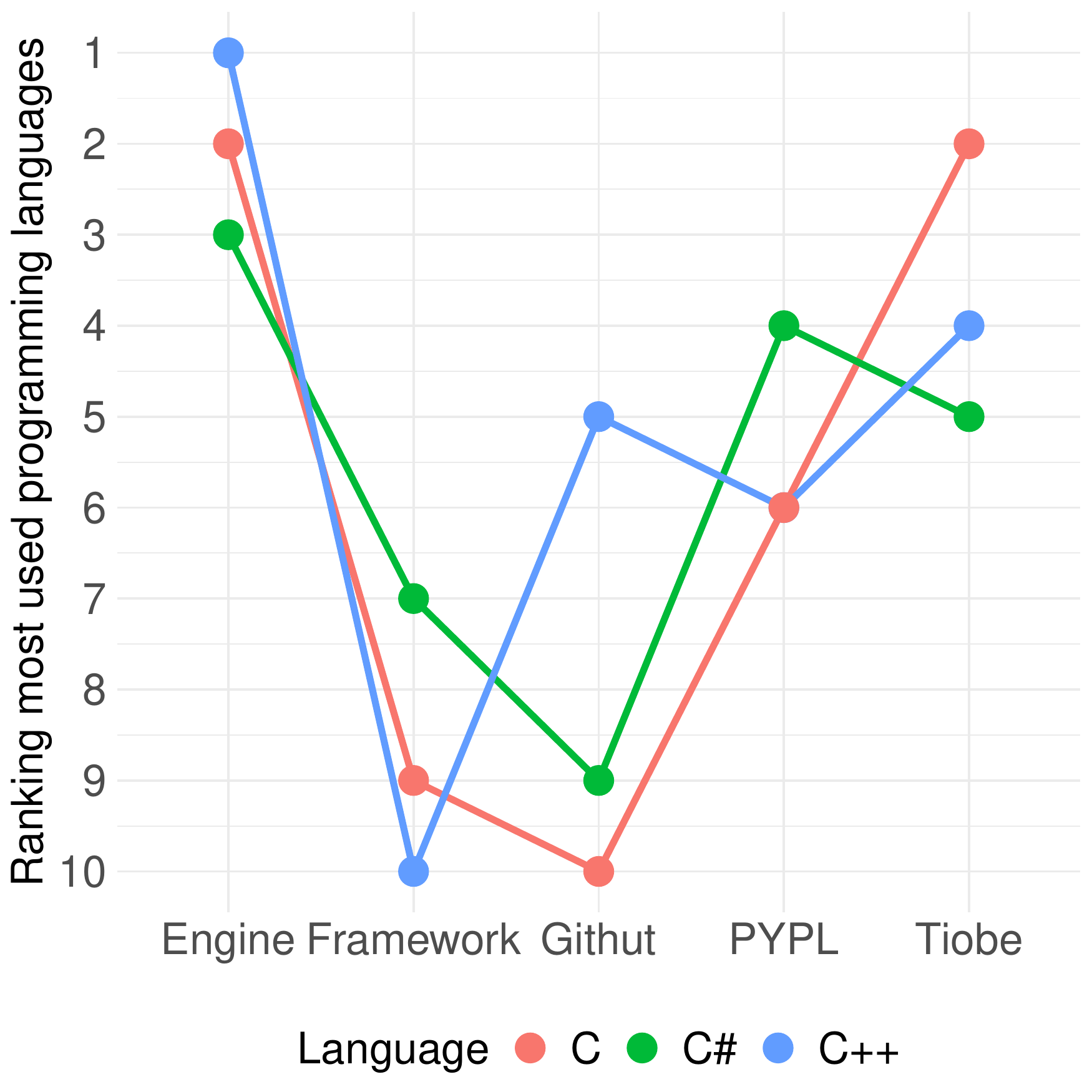}
\caption{Most used languages in Game Engines.}
\label{fig:rank-lang-lowlevel}
\end{subfigure}

\begin{subfigure}{1\linewidth}
\centering
\includegraphics[width=.9\linewidth]{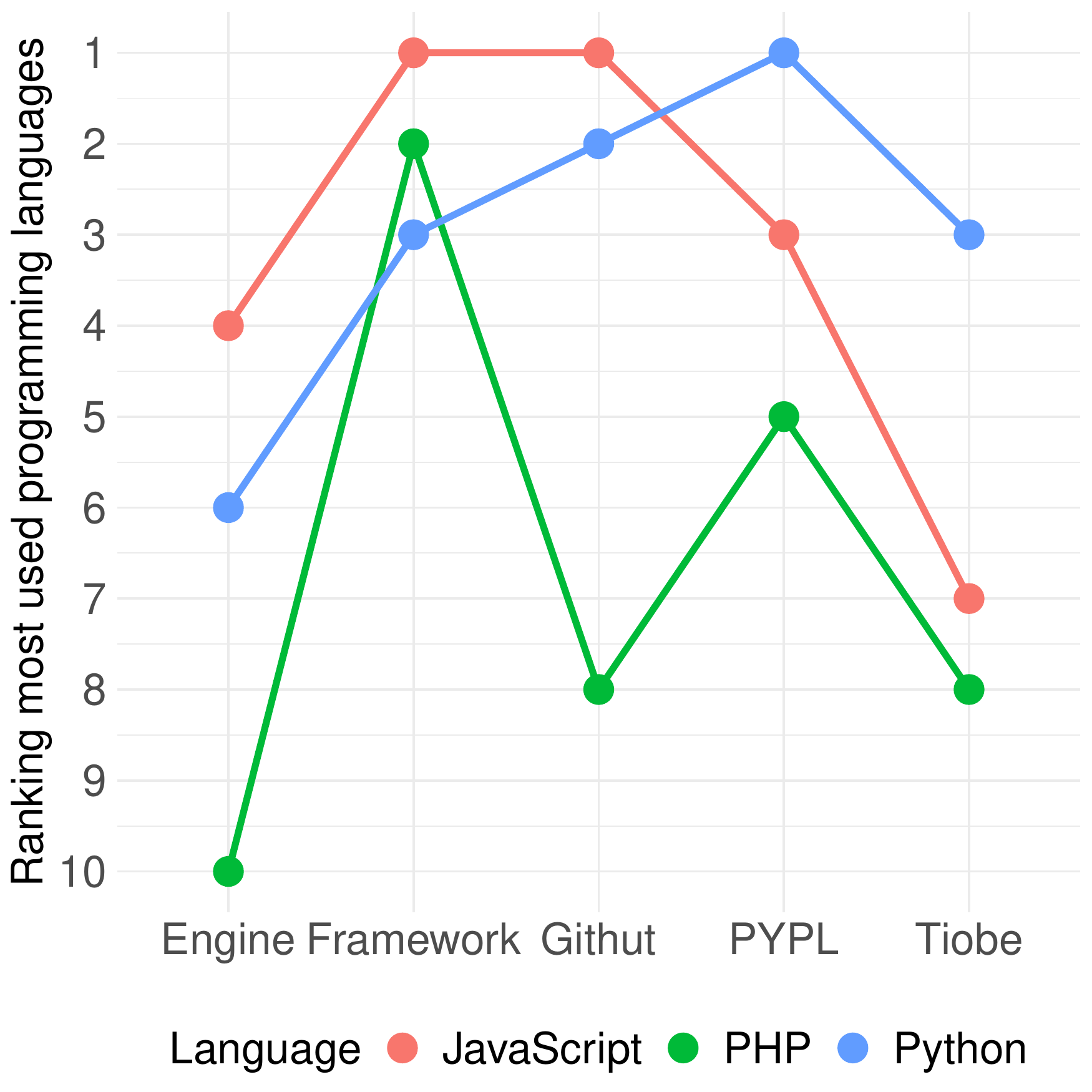}
\caption{Most used languages in Frameworks.}
\label{fig:rank-lang-scripting}
\end{subfigure}
\caption{Ranking of the languages used in engines and frameworks compared to their global uses.}
\label{fig:rank-lang}
\end{figure}

\autoref{fig:rank-lang} shows the ranking of the top six most used languages in engines and frameworks and compare it to other three global sources of programming languages usage in 2019: GitHut 2.0\footnote{\url{https://madnight.github.io/githut/}} (Third Quarter), which uses the GitHub API to query the most used languages in the public repositories, Tiobe index\footnote{\url{https://www.tiobe.com/tiobe-index/}} (November), which uses a set of metrics together with results from search engines, and PYPL\footnote{\url{https://pypl.github.io/PYPL.html}} (PopularitY of Programming Language, December), which is a ranking created by analysing how often language tutorials are searched on Google.

C++ is the most popular language for engines but is only the 10\textsuperscript{th} most popular in frameworks, 5\textsuperscript{th} in GitHut, 4\textsuperscript{th} in Tiobe, and 6\textsuperscript{th} in PYPL. C is used in engines and is in 2\textsuperscript{nd} position in Tiobe but not so popular according to the other sources. JavaScript is embraced by the open-source community and received lots of attention in searches but Tiobe puts it in the 7\textsuperscript{th} place. The popularity of the programming languages in engines is more aligned to the Tiobe index than to the other sources. In contrast, popular languages in frameworks are more aligned to GitHut and PYPL rankings. Game engines are more aligned with the commercial market and less with open-source projects.

\subsection*{\rqBb}

\autoref{tab:licenses} shows the top 10 most used licenses. Differently from the languages, the distribution of licenses is similar between engines and frameworks. The MIT License is most used for both types of projects with 46\%. ``Other'' licenses (not reported by GitHub) are the second most popular with 18\%. 9\% of the projects do not have an explicit license. The remaining ones form 27\%.

\begin{table}[width=1\linewidth, pos=htb!]
\caption{Most Used Licenses.}
\label{tab:licenses}
\centering
\begin{tabular*}{\tblwidth}{@{}Lrrrrrr@{}}
\toprule
 & \multicolumn{2}{l}{Engine} & \multicolumn{2}{l}{Framework} & \multicolumn{2}{l}{Total} \\ \cmidrule(l){2-7}
 & \multicolumn{1}{l}{N} & \multicolumn{1}{l}{\%} & \multicolumn{1}{l}{N} & \multicolumn{1}{l}{\%} & \multicolumn{1}{l}{N} & \multicolumn{1}{l}{\%} \\ \midrule
MIT License & 116 & 41\% & 142 & 50\% & 258 & 46\% \\
Other & 55 & 20\% & 48 & 17\% & 103 & 18\% \\
\emph{No licence specified} & 30 & 11\% & 22 & 8\% & 52 & 9\% \\
Apache License 2.0 & 17 & 6\% & 35 & 12\% & 52 & 9\% \\
GNU GPL v3.0 & 28 & 10\% & 10 & 4\% & 38 & 7\% \\
GNU LGPL v3.0 & 8 & 3\% & 5 & 2\% & 13 & 2\% \\
BSD 3 & 2 & 1\% & 7 & 2\% & 9 & 2\% \\
GNU GPL v2.0 & 7 & 2\% & 2 & 1\% & 9 & 2\% \\
zlib License & 6 & 2\% & 0 & 0\% & 6 & 1\% \\
GNU AGPL v3.0 & 2 & 1\% & 3 & 1\% & 5 & 1\% \\
BSD 2 & 3 & 1\% & 1 & 0\% & 4 & 1\% \\
GNU LGPL v2.1 & 1 & 0\% & 3 & 1\% & 4 & 1\% \\
Mozilla PL 2.0 & 2 & 1\% & 1 & 0\% & 3 & 1\% \\
The Unlicense & 2 & 1\% & 1 & 0\% & 3 & 1\% \\
Artistic License 2.0 & 1 & 0\% & 0 & 0\% & 1 & 0\% \\
Boost SL 1.0 & 0 & 0\% & 1 & 0\% & 1 & 0\% \\
CC Attribution 4.0 & 1 & 0\% & 0 & 0\% & 1 & 0\% \\
Eclipse PL 1.0 & 0 & 0\% & 1 & 0\% & 1 & 0\% \\
Microsoft PL & 1 & 0\% & 0 & 0\% & 1 & 0\% \\ \bottomrule
\end{tabular*}
\end{table}

According to GitHut 2.0, the licenses in GitHub project are ranked as follows: MIT 54\%, Apache 16\%, GPL 2 13\%, GPL 3 10\%, and BSD 3 5\%, which is similar to the rankings in engines and frameworks. 
Projects with ``Other'' licenses are a big part of this data, which, according to \citet{Vendome2017}, are prone to migrate towards Apache or GPL licenses.
Finally, the MIT license is popular thanks to its permissive model, which fits well with most open-source projects.

The licenses might only apply to the engines or frameworks and \emph{not} to the games or software. For example, games created with Godot have their creators as sole copyright owners but must include its license:

\begin{quote}
\textit{``Godot Engine's license terms and copyright do not apply to the content you create with it; you are free to license your games how you see best fit, and will be their sole copyright owner(s). Note however that the Godot Engine binary that you would distribute with your game is a copy of the `Software' as defined in the license, and you are therefore required to include the copyright notice and license statement somewhere in your documentation.''}\newline -- \url{https://godotengine.org/license}
\end{quote}

\subsection*{\rqBc}

\begin{figure}[pos=htb!]
\centering
\begin{subfigure}{.32\linewidth}
\includegraphics[width=1\linewidth]{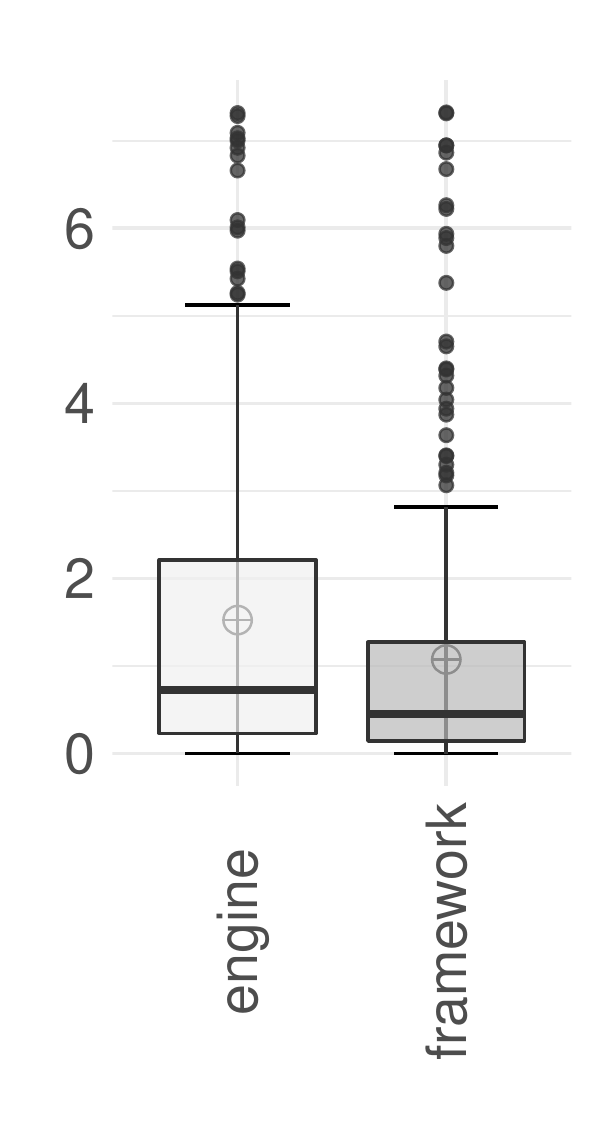}
\caption{Language size}
\end{subfigure}
\begin{subfigure}{.32\linewidth}
\includegraphics[width=1\linewidth]{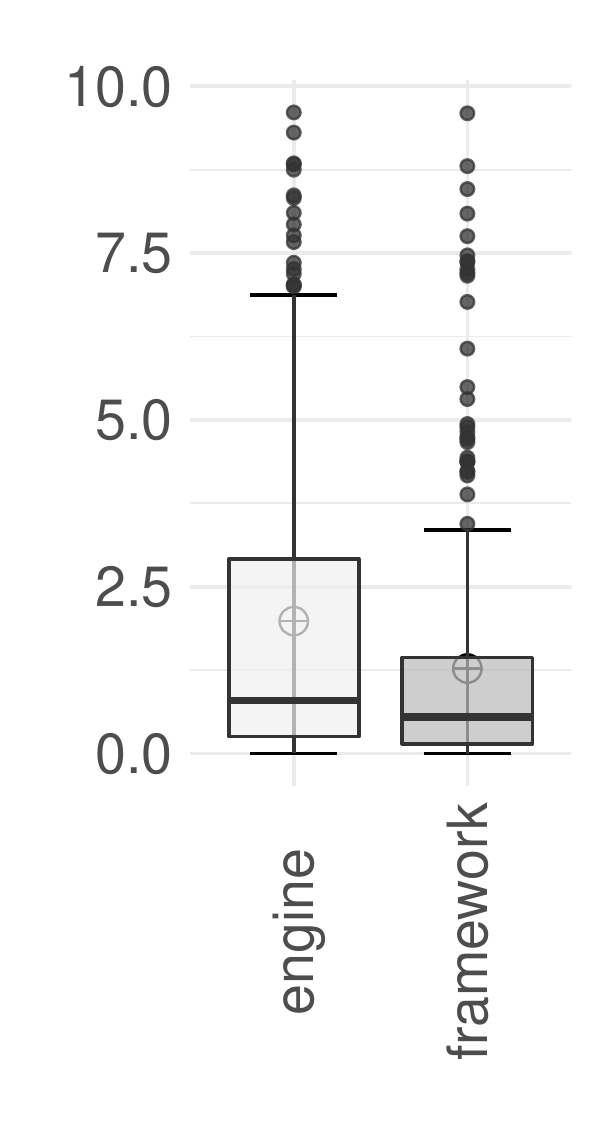}
\caption{Total size}
\end{subfigure}
\begin{subfigure}{.32\linewidth}
\includegraphics[width=1\linewidth]{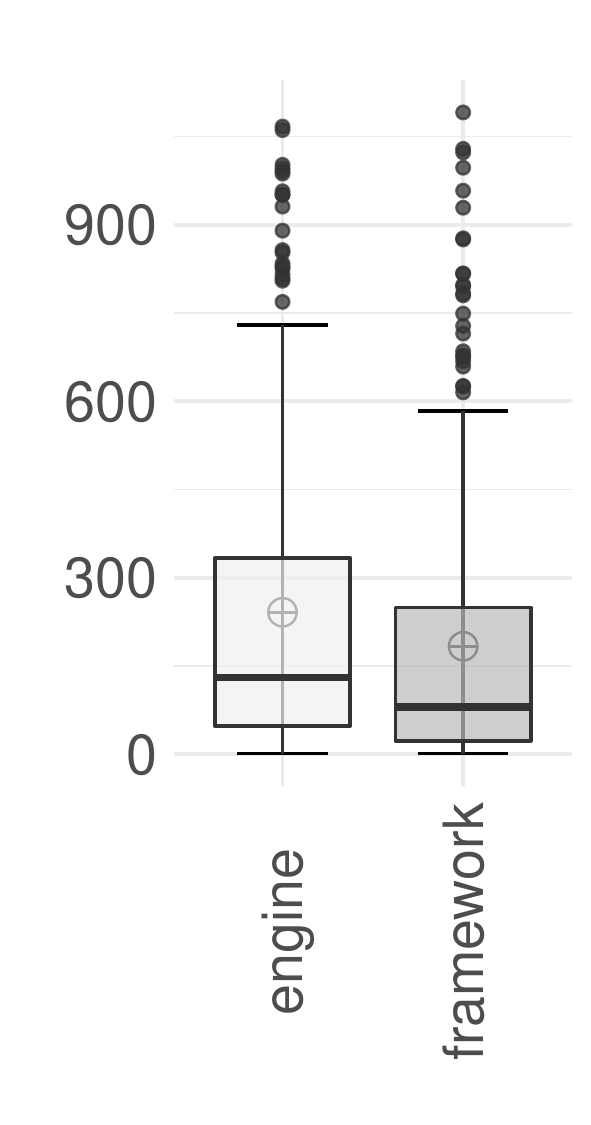}
\caption{Number of files}
\end{subfigure}
\caption{Boxplots -- \rqBc \\
(a) \mBcA \\
(b) \mBcB \\
(c) \mBcC
}
\label{fig:boxplot-rq2.3}
\end{figure}

We considered \emph{main\_language\_size}, \emph{total\_size}, and \emph{n\_files}. They show larger values for engines when compared to frameworks. Considering the medians, engines have around 50\% higher median values regarding size of the main language (1.09MB), total size of the project (1.22MB), and number of files (171 files). The boxplots in \autoref{fig:boxplot-rq2.3} help to identify the differences among variables: game engines are larger than frameworks, on average.

\subsection*{\rqBd}

\begin{figure}[pos=htb!]
\centering
\begin{subfigure}{.32\linewidth}
\includegraphics[width=1\linewidth]{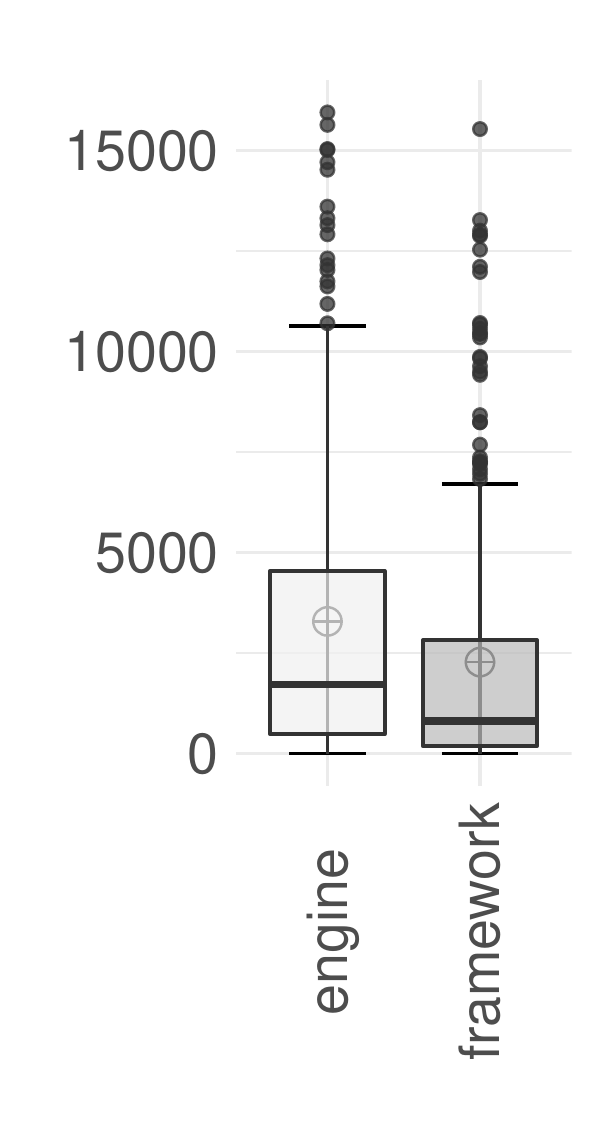}
\caption{Functions}
\end{subfigure}
\begin{subfigure}{.32\linewidth}
\includegraphics[width=1\linewidth]{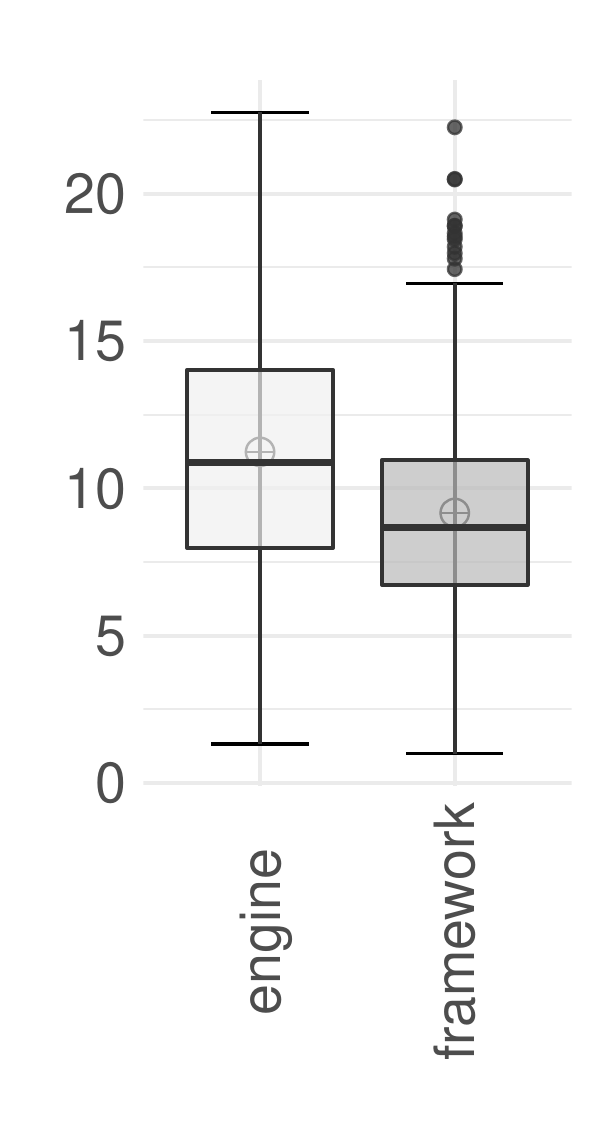}
\caption{Code lines}
\end{subfigure}
\begin{subfigure}{.32\linewidth}
\includegraphics[width=1\linewidth]{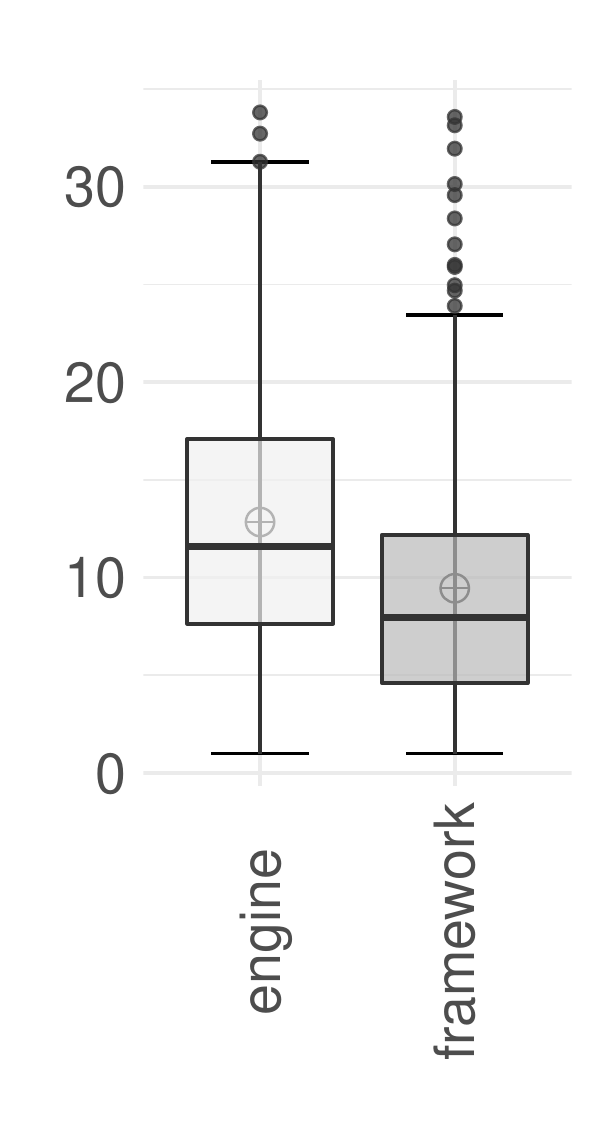}
\caption{Functions per file}
\end{subfigure}
\caption{Boxplots -- \rqBd \\
(a) \mBdA \\
(b) \mBdB \\
(c) \mBdC}
\label{fig:boxplot-rq2.4}
\end{figure}

We considered \emph{n\_func}, \emph{nloc\_mean}, and \emph{func\_per\_file\_mean}. The boxplots in \autoref{fig:boxplot-rq2.4} show that engines have larger values when compared to frameworks. Considering medians, engines have around 30\% more functions per file and 20\% more functions and lines of code per function.

\subsection*{\rqBe}

\begin{figure}[pos=htb!]
\includegraphics[width=1\linewidth]{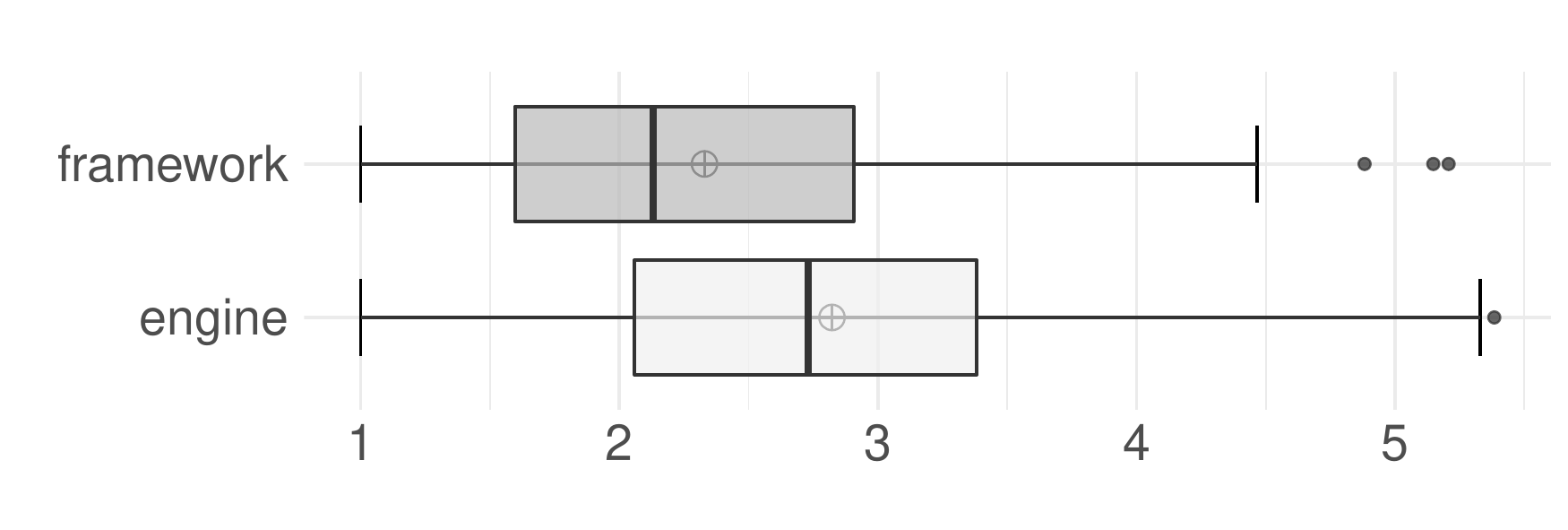}
\caption{Boxplot of the Cyclomatic Complexity (CC).}
\label{fig:boxplot-rq2.5}
\end{figure}

We considered \emph{cc\_mean} to assess functions complexities. The median for engines is about 23\% greater than the frameworks, which correspond to a complexity of less than 1. \autoref{fig:boxplot-rq2.5} illustrates this difference. We also identified 16 projects (10 engines and 6 frameworks) with median complexities greater than 5.

Factoring out low-level languages, e.g., C++, the complexity of the functions are similar between engines and frameworks. The number of files and average functions per file are similar. The total number of functions, however, is different: engines have twice as many functions (median values).

Programming languages vary greatly, as the game engines are written mostly in compiled languages, while frameworks in interpreted ones. Both types of projects prefer the MIT license, very suitable to open-source projects. Although game engines are, on average, bigger and more complex than frameworks, this difference is small.

\clearpage

\section{Detailed Results for \rqC}
\label{sec:rq3-historic}

\begin{table*}[width=1\textwidth, pos=tb!]
\caption{Descriptive Statistics: \rqC. Normality \textless 0.01 means the data is not normally distributed.}
\label{tab:ds-rq3}
\centering
\begin{tabular*}{\tblwidth}{@{}lLlrrrrrr@{}}
\toprule
RQs                    & Variable       & Type      & Mean & Std.Dev. & Median & Min & Max & Normality \\ \midrule
\multirow{2}{*}{RQ3.1} & tags\_releases\_count & engine  & 15.82 & 52.20 & 1.00 & 0.00 & 657.00 & \textless 0.01 \\
 & tags\_releases\_count & framework & 82.24 & 216.37 & 32.00 & 0.00 & 2,678.00 & \textless 0.01 \\ \midrule
\multirow{2}{*}{RQ3.2} & lifespan (weeks) & engine  & 155.70 & 113.39 & 135.79 & 0.00 & 530.43 & \textless 0.01 \\
& lifespan (weeks) & framework & 215.30 & 129.79 & 182.14 & 5.71 & 590.71 & \textless 0.01 \\ \midrule
\multirow{4}{*}{RQ3.3} & commits\_count & engine  & 2,029.93 & 4,553.92 & 616.00 & 7.00 & 37,026.00 & \textless 0.01 \\
& commits\_count & framework & 3,463.88 & 8,581.04 & 833.50 & 20.00 & 87,774.00 & \textless 0.01 \\ \addlinespace
& commits\_per\_time & engine  & 3.44 & 7.71 & 1.04 & 0.01 & 62.68 & \textless 0.01 \\
& commits\_per\_time & framework & 5.86 & 14.53 & 1.41 & 0.03 & 148.59 & \textless 0.01 \\ \midrule
\multirow{6}{*}{RQ3.4} & lines\_added   & engine    & 2,403.59 & 9,179.15 & 424.20 & 7.00  & 94,597.77 & \textless 0.01 \\
                       & lines\_added   & framework & 776.62   & 2,750.07 & 169.46 & 5.80  & 26,099.63 & \textless 0.01 \\ \addlinespace
                       & lines\_removed & engine    & 644.34   & 1,343.49 & 175.44 & 0.00  & 11,957.13 & \textless 0.01 \\
                       & lines\_removed & framework & 434.67   & 1,421.13 & 104.84 & 2.33  & 15,450.48 & \textless 0.01 \\
                       \addlinespace
                       & code\_churn    & engine    & 3,074.94 & 9,512.87 & 423.86 & 7.00  & 95,426.69 & \textless 0.01 \\
                       & code\_churn    & framework & 1,211.28 & 4,015.20 & 163.79 & 10.67 & 41,550.10 & \textless 0.01 \\
\bottomrule
\end{tabular*}
\end{table*}

\subsection*{\rqCa}

\begin{figure}[pos=htb!]
\centering
\includegraphics[width=1\linewidth]{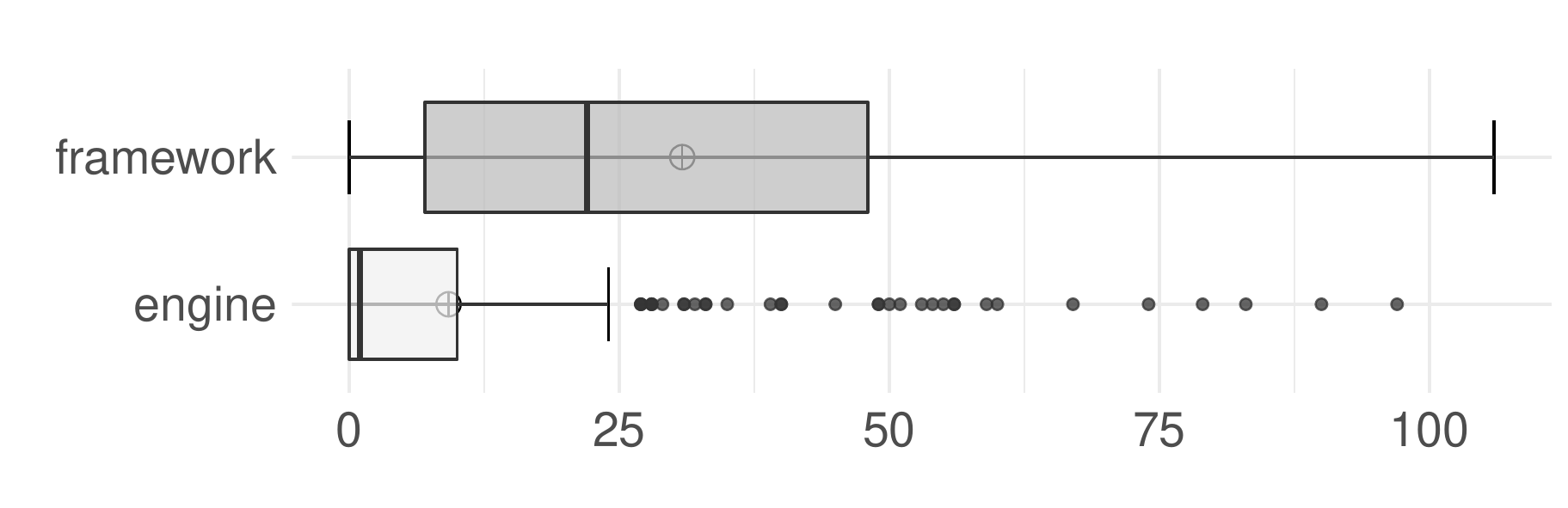}
\caption{Boxplot of number of tags -- \rqCa}
\label{fig:boxplot-rq3.1}
\end{figure}

Around 40\% of the engines (112 projects) do not have any tag. Only 8\% of the frameworks (23 projects) are lacking them. Most engines have between 0 and 11 tags while frameworks have between 9 to 88. Frameworks release new versions more often. \autoref{fig:boxplot-rq3.1} shows the boxplots for the numbers of tags. The dots represent the outliers as the majority of the engines have zero or few tags.

\subsection*{\rqCb}

\begin{figure}[pos=htb!]
\centering
\includegraphics[width=1\linewidth]{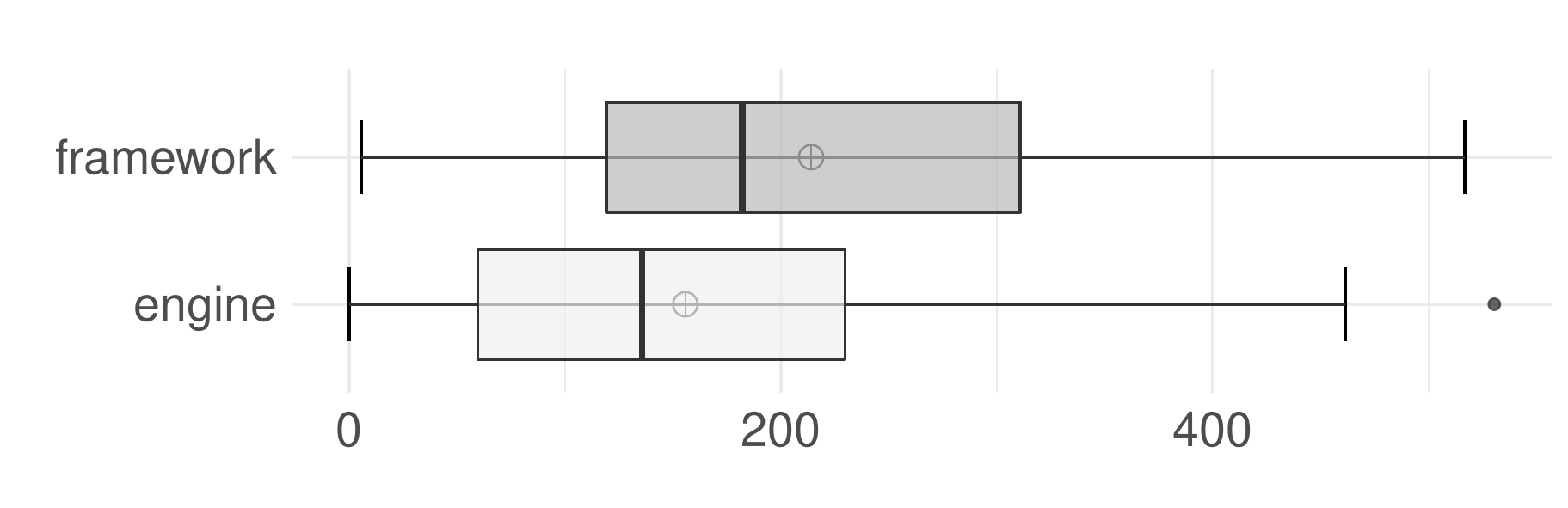}
\caption{Boxplot of lifespan in weeks -- \rqCb}
\label{fig:boxplot-rq3.2}
\end{figure}

\autoref{fig:boxplot-rq3.2} shows the distributions of engines and frameworks lifetimes in weeks: both have similar shapes, with more projects in the last years. Considering median values, engines and frameworks are 2.6 and 3.5 years-old, respectively. Open-source engines are more recent when compared to open-source frameworks.

\subsection*{\rqCc}

\begin{figure}[pos=htb!]
\centering
\begin{subfigure}{.35\linewidth}
\includegraphics[width=1\linewidth]{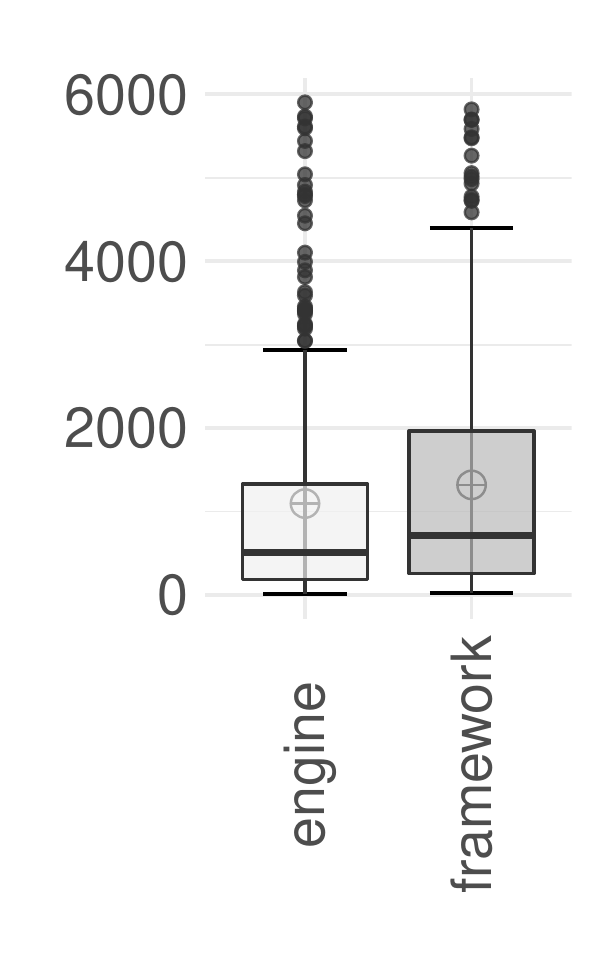}
\caption{Commits}
\label{fig:boxplot-goalB-commits-count}
\end{subfigure}
\quad
\begin{subfigure}{.35\linewidth}
\includegraphics[width=1\linewidth]{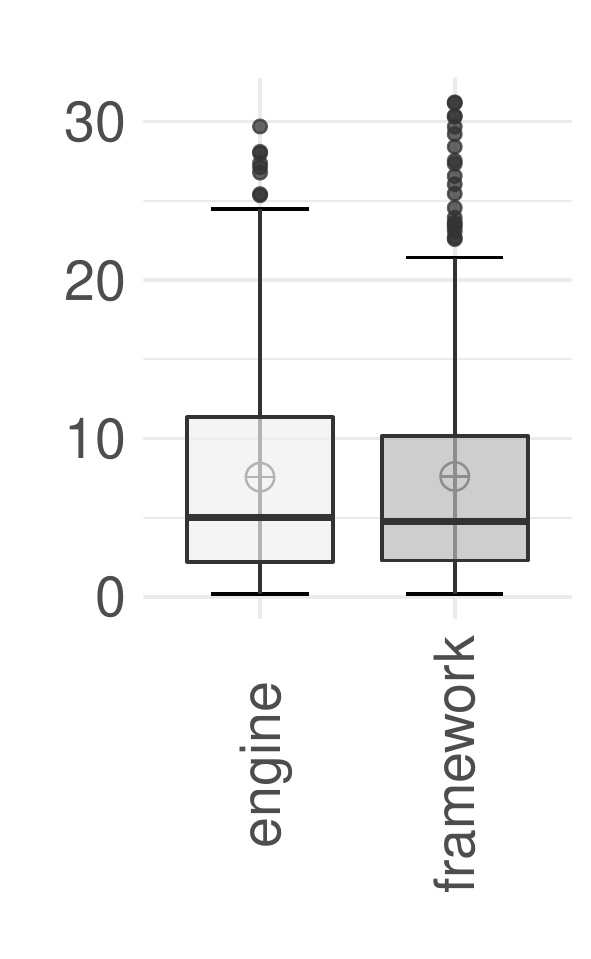}
\caption{Commits per week}
\label{fig:boxplot-goalB-commits-time}
\end{subfigure}
\caption{Boxplots -- \rqCc}
\label{fig:boxplot-goalB-commits}
\end{figure}

\autoref{fig:boxplot-goalB-commits-count} presents the distribution for \emph{commits\_count}. The frequency and number of commits is larger for frameworks in total. Most engines have more than 616 commits, while frameworks have more than 833, in the median.

\autoref{fig:boxplot-goalB-commits-time} depicts the distribution for \emph{commits\_per\_time}, which represents the number of commits (\emph{commits\_count}) averaged by the projects' lifetime (\emph{lifespan}). Interestingly, engines are more active than frameworks when considering the number of commits overtime. In fact, 47\% of the engines have at least one commit per week, while this activity is achieved by 40\% of frameworks. This behavior---together with the results presented in RQ2.2---may indicate that engines are less mature than frameworks.

\subsection*{\rqCd}

To answer this research question we investigated three different metrics: \emph{lines\_added}, \emph{lines\_removed}, and \emph{code\_churn}. To keep our comparison at project level, we averaged the values of these metrics based on the number of commits each project has. Overall, engines added and removed more lines than frameworks. In median basis, 424.20 lines were added by engines, against 169.46 lines for frameworks. When it comes to lines removed, engines deleted 175.44 of them, while frameworks removed 104.84; also median values. This results in a code churn that is 2.6x times higher for engines when compared to frameworks (423.86 vs 163.79), which means that engines are more likely to have major changes than frameworks.

\clearpage

\section{Detailed Results for \rqD}
\label{sec:rq4-community}

\begin{table*}[width=1\textwidth, pos=tb!]
\caption{Descriptive Statistics, \rqD. Normality \textless 0.01 means the data is not normally distributed.}
\label{tab:ds-rq4}
\centering
\begin{tabular*}{\tblwidth}{@{}lLlrrrrrr@{}}
\toprule
RQs & Variable & Type & Mean & Std.Dev. & Median & Min & Max & Normality \\ \midrule
\multirow{2}{*}{RQ4.1} & truck\_factor & engine & 1.32 & 0.91 & 1 & 1 & 8 & \textless 0.01 \\
 & truck\_factor & framework & 1.58 & 1.88 & 1 & 1 & 25 & \textless 0.01 \\ \midrule
\multirow{4}{*}{RQ4.2} & stargazers\_count & engine  & 659.45 & 2,140.45 & 44.5 & 2 & 23775 & \textless 0.01 \\
 & stargazers\_count & framework & 4,017.36 & 11,671.63 & 556.5 & 111 & 145516 & \textless 0.01 \\ \addlinespace
 & contributors\_count & engine  & 18.95 & 52.66 & 3 & 2 & 435 & \textless 0.01 \\
 & contributors\_count & framework & 57.09 & 94.34 & 15 & 2 & 403 & \textless 0.01 \\ \midrule
 \multirow{6}{*}{RQ4.3} & issues\_count & engine  & 494.19 & 2,453.87 & 44 & 0 & 30,317 & \textless 0.01\\
 & issues\_count & framework & 1,534.70 & 4,075.97 & 254 & 3 & 36,757 & \textless 0.01\\ \addlinespace
 & closed\_issues\_count & engine & 428.5 & 2,122.72 & 34.5 & 0 & 24,861 & \textless 0.01\\
 & closed\_issues\_count & framework & 1,449.18 & 3,496.88 & 215.5 & 0 & 35,659 & \textless 0.01\\ \addlinespace
 & closed\_issues\_rate & engine & 80\% & 20\% & 85\% & 0\% & 100\% & \textless 0.01\\
 & closed\_issues\_rate & framework & 87\% & 14\% & 91\% & 0\% & 100\% & \textless 0.01\\
\bottomrule
\end{tabular*}
\end{table*}

\subsection*{\rqDa}

\autoref{tab:truck} shows the truck-factor values and numbers of contributors per project. The distribution of the truck-factor between engines and frameworks are similar with the majority of the projects having a value equal to one (82\% for engines and 73\% for frameworks). The engine with the highest truck-factor is PGZero with value of 8. Three frameworks have truck-factor values higher than 8: Django (9), Rails (13), and FrameworkBenchmarks (25). As a comparison, Linux\footnote{\url{https://github.com/torvalds/linux}} has a truck-factor of 57 and Git of 12 \cite{Avelino2016}.

\begin{table}[width=1\textwidth, pos=htb!]
\centering
\caption{Truck-factor values and medians of contributors.}
\label{tab:truck}
\begin{tabular}{@{}crrrr@{}}
\toprule
\multicolumn{1}{l}{} & \multicolumn{2}{l}{Frameworks} & \multicolumn{2}{l}{Engines} \\ \cmidrule(l){2-5}
\multicolumn{1}{l}{Truck-factor} & \multicolumn{1}{c}{N} & \multicolumn{1}{c}{Contributors} & \multicolumn{1}{c}{N} & \multicolumn{1}{c}{Contributors} \\ \midrule
1 & 208 & 11 & 231 & 3 \\
2 & 46 & 46 & 35 & 13 \\
3 & 10 & 75.5 & 7 & 47 \\
4 & 11 & 75 & 4 & 50 \\
5 & 1 & 355 & 1 & 216 \\
6 & 2 & 299 & 1 & 312 \\
7 & -- & -- & 2 & 294 \\
8 & 1 & 69 & 1 & 32 \\
9 & 1 & 403 & -- & -- \\
13 & 1 & 377 & -- & -- \\
25 & 1 & 374 & -- & -- \\ \bottomrule
\end{tabular}
\end{table}

The median of contributors follows an direct relation: the higher the truck-factor, the higher the number of contributors. The exceptions are the engine PGZero (Python) with 32 contributors and the framework Sofa (C++) with 69 contributors, both with truck-factor 8.

\subsection*{\rqDb}

\begin{figure}[pos=htb!]
\begin{subfigure}{1\linewidth}
\includegraphics[width=1\linewidth]{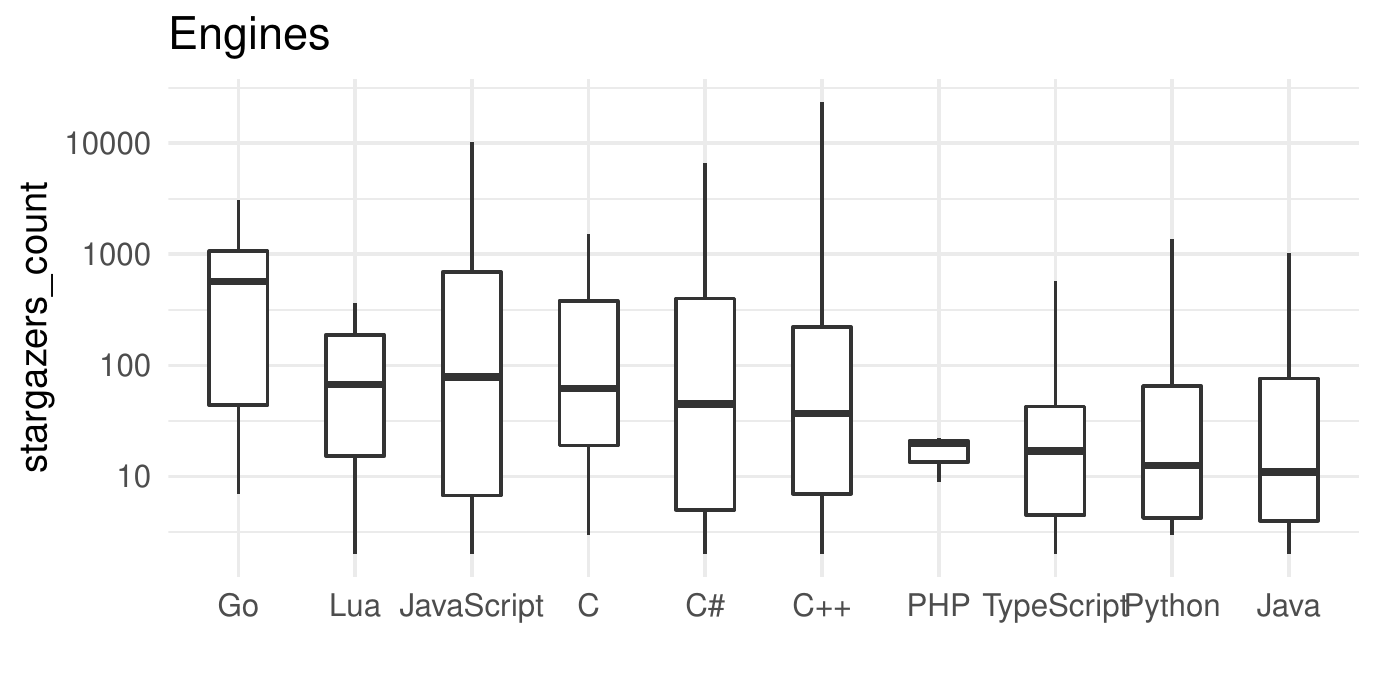}
\label{fig:pop-eng}
\end{subfigure}

\begin{subfigure}{1\linewidth}
\includegraphics[width=1\linewidth]{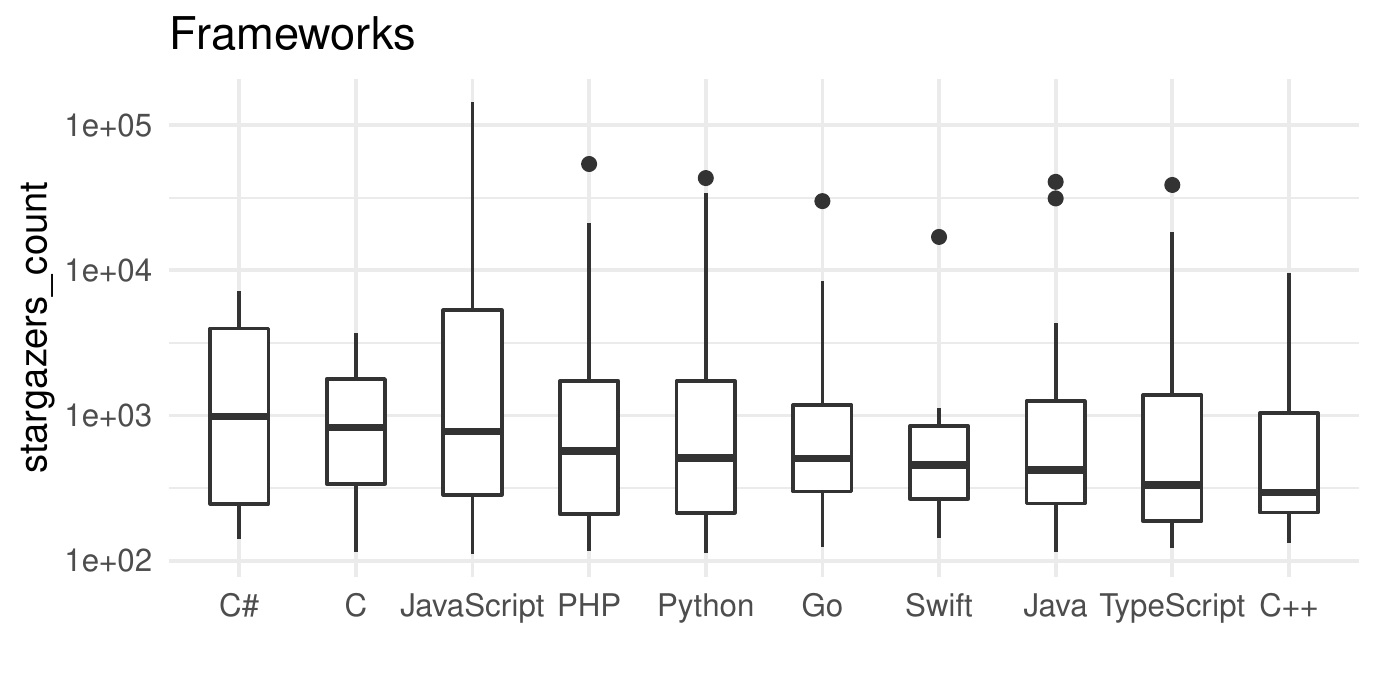}
\label{fig:pop-fra}
\end{subfigure}

\caption{Popularity of the Top 10 Most Used Languages for Engines and Frameworks Ordered by Medians.}
\label{fig:popularity}
\end{figure}

\autoref{fig:popularity} shows the popularity of the projects considering the top 10 most used languages (\autoref{tab:languages}) ordered by median numbers of stars. Engines written in Go have the highest popularity although they are only 14. JavaScript is the second most popular language followed by the C family. Although C++ makes up the majority of the engines, it is only the fifth most popular. C\# is the most popular language for frameworks, but with only 15 projects. JavaScript and C are second and third, respectively.

\subsection*{\rqDc}

As observed in \autoref{tab:ds-rq4}, issues activity on frameworks are higher when compared to engines. For instance, 50\% of the frameworks have at least 254 issues reported (i.e., median). This number drops to 44 for engines. The number of closed issues present a similar difference: half of the projects have at least 215 and 34 issues closed for frameworks and engines, respectively. When it comes to the rate of closed issues, both kinds of systems present similar results, though. In median, engines have closed 85\% of the issues reported so far, while frameworks have closed 91\% of them.

\end{document}